\documentclass[11pt]{article}
\usepackage{amssymb}
\usepackage{mathtools}
\usepackage{epsf}
\usepackage{caption}
\usepackage{amsmath}
\usepackage{amsthm}
\newcommand{\sJM}{\mathop{\sum\nolimits\sp{\ne}}}

\newcommand{\oR}{{\mathbb R}}
\newcommand{\oE}{{\mathbb E}}

\newtheorem{algo}{Algorithm}

\begin{document}
\thispagestyle{empty}

\begin{center}
{\bf \large Exploring seismic hazard in the Groningen gas field \\
using adaptive kernel smoothing and inhomogeneous summary statistics}\\[0.5in]

M.N.M.\ van Lieshout$\mbox{}^{a,b}$ and Z.\ Baki$\mbox{}^{b,a}$\\[0.2in]

$\mbox{}^a$
CWI, P.O.~Box 94079, NL-1090 GB Amsterdam, The Netherlands 

\

$\mbox{}^b$
Faculty of Electrical Engineering, Mathematics and Computer Science, 
University of Twente, P.O.~Box 217, NL-7500 AE, Enschede, The Netherlands.
\end{center}

\bigskip

\noindent
ABSTRACT: The discovery of gas in Groningen in 1959 has been a massive boon
to the Dutch economy. From the 1990s onwards though, gas production has led 
to induced seismicity. In this paper, we carry out a comprehensive exploratory 
analysis of the spatio-temporal earthquake catalogue. We develop a non-parametric 
adaptive kernel smoothing technique to estimate the spatio-temporal hazard map 
and to interpolate monthly well-based gas production statistics. Second- and 
higher-order inhomogeneous summary statistics are used to show that the state 
of the art rate-and-state models for the prediction of seismic hazard 
fail to capture inter-event interaction in the earthquake catalogue. Based
on these findings, we suggest a modified rate-and-state model that also takes 
into account changes in gas production volumes and uncertainty in the pore 
pressure field.

\noindent
{\bf Keywords:} adaptive bandwidth selection, induced seismicity,
inhomogeneous summary statistics, kernel smoothing, pore pressure,
spatio-temporal point process.


\section{Introduction}

In 1959, a large gas field was discovered in Groningen, a province in
the North of The Netherlands. Its recoverable gas volume has been estimated
at around $2,900$ billion Normal cubic metres (Nbcm). The extraction
rate has varied considerably over the years. After a modest start, large
amounts were being produced annually during the early 1970s rising to
about $85$ Nbcm in 1976. During the next decade, the production volumes
tended to decrease, followed by somewhat higher values during the first half
of the 1990s. Production fell again during the second half of the decade,
before rising in the new millennium to over $53$ bcm in 2013. From the
1990s earthquakes were being registered in the previously tectonically
inactive Groningen region. Especially the one near Huizinge in August
2012 with a magnitude of $3.6$ led to a public demand for a reduction
of gas production volumes. The government reacted with legislation to
phase out gas extraction and, by 2020, production had fallen to less
than $8$ Nbcm.  

Numerous studies on the Groningen field have been conducted. For example,
Geerdink \cite{Geer14} models the times in between earthquakes in terms of the
cumulative and annual production rates, pressure, subsidence and fault
zones. More recent examples of such a study include Post et al.\ \cite{Post21} 
and Trampert et al.\ \cite{Tram22}.
Van Hove et al.\ \cite{Hove15} propose a Poisson auto-regression model
for the annual hazard maps in terms of subsidence, fault lines and gas
extraction in previous years. Both Hettema et al.\ \cite{Hett17} and
Vlek  \cite{Vlek19} explore
the temporal development of seismicity in Groningen by proposing a linear
model for the relation between the number of earthquakes over specific
periods and gas production volumes. Sijacic et al.\ \cite{Sija17}
focus on the detection of changes in the rate of a temporal Poisson point
process by Bayesian and frequentist methods. Bourne et al.\ \cite{Bour18}
modify Ogata's space-time model \cite{Ogat88} to include changes in stress
level and estimate the probability of fault failures. Other papers,
notably Candela et al.\ \cite{Cand19}, Dempsey and Suckale \cite{DempSuck17}
and Richter et al.\ \cite{Rich20}, discuss the modelling of seismicity in
relation to stress changes based on a differential equation and embed
these in a space-time Poisson point process.

In a previous paper \cite{BakiLies22}, we investigated the temporal
development of seismicity in Groningen including data up to 2020 using
cumulative and recent gas production as dependent variables in a 
regression model. We concluded that a decrease in production leads to
decreased seismicity. Here, we extend the analysis to take into account
spatial variations. First, we calculate non-parametric estimates for the
spatio-temporal hazard map by means of an adaptive kernel smoother
(cf.\ Abramson \cite{Abra82a}, Davies et al.\ \cite{Davi18} and Van Lieshout 
\cite{Lies21}) and generalise the bandwidth selection approach
suggested by Van Lieshout \cite{Lies21} to the space-time domain. 
Using this map, we investigate whether a Poisson point process model
would suffice. Employing inhomogeneous summary statistics, we find that 
there is interaction which cannot be accounted for by Poisson models, 
including the state of the art rate-and-state models in Candela et al.\ 
\cite{Cand19}, Dempsey and Suckale \cite{DempSuck17} and Richter et al.\ 
\cite{Rich20}. Since rate-and-state models rely on differential equations 
for changes in Coulomb stress or, equivalently, pore pressure, we shift 
our attention to pressure and production data in the public domain. The 
production values are measured monthly at wells whereas pressure is gauged 
at irregular times at some wells as well as at several observation and 
seismic monitoring stations. To obtain a gas production map, adaptive 
kernel smoothing applies. Since mass must be preserved, Van Lieshout's 
local edge correction \cite{Lies12} is required. For the pressure values,
we fit a Gaussian random field, the mean function of which is modelled as a 
polynomial in space and time. Finally, we propose a modification of the
rate-and-state models of Candela et al.\ \cite{Cand19}, Dempsey and Suckale 
\cite{DempSuck17} and Richter et al.\ \cite{Rich20} that exhibits
clustering, accounts for the uncertainty in pore pressure, takes into
account the varying gas production, and is amenable to monitoring by
means of Markov chain Monte Carlo methods based on the history of
recorded earthquakes.

The plan of this paper is as follows. In Section~\ref{S:data}, we
describe the data. Section~\ref{S:explore} carries out a comprehensive
second-order analysis, Section~\ref{S:covariates} is devoted to
extrapolation of gas production and pore pressure measurements from
wells to field. The paper closes with our proposed modification of the 
Coulomb rate-and-state seismicity model.

\section{Data}
\label{S:data}

Data on the Groningen gas field and the induced earthquakes is available 
at various sources.

\begin{figure}[hbt]
\centering
\centerline{
\epsfxsize=0.45\hsize
\epsffile{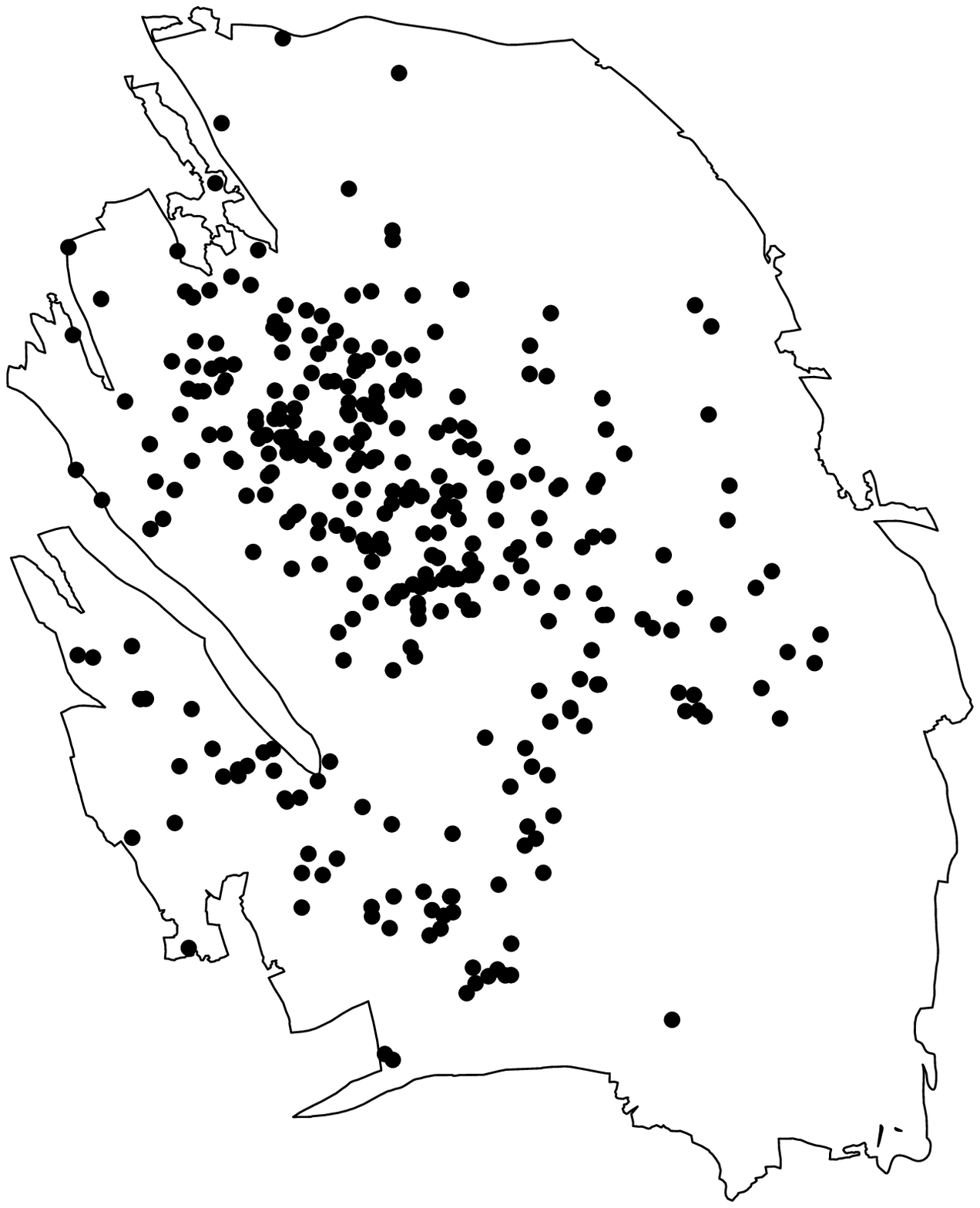}
\epsfxsize=0.45\hsize
\epsffile{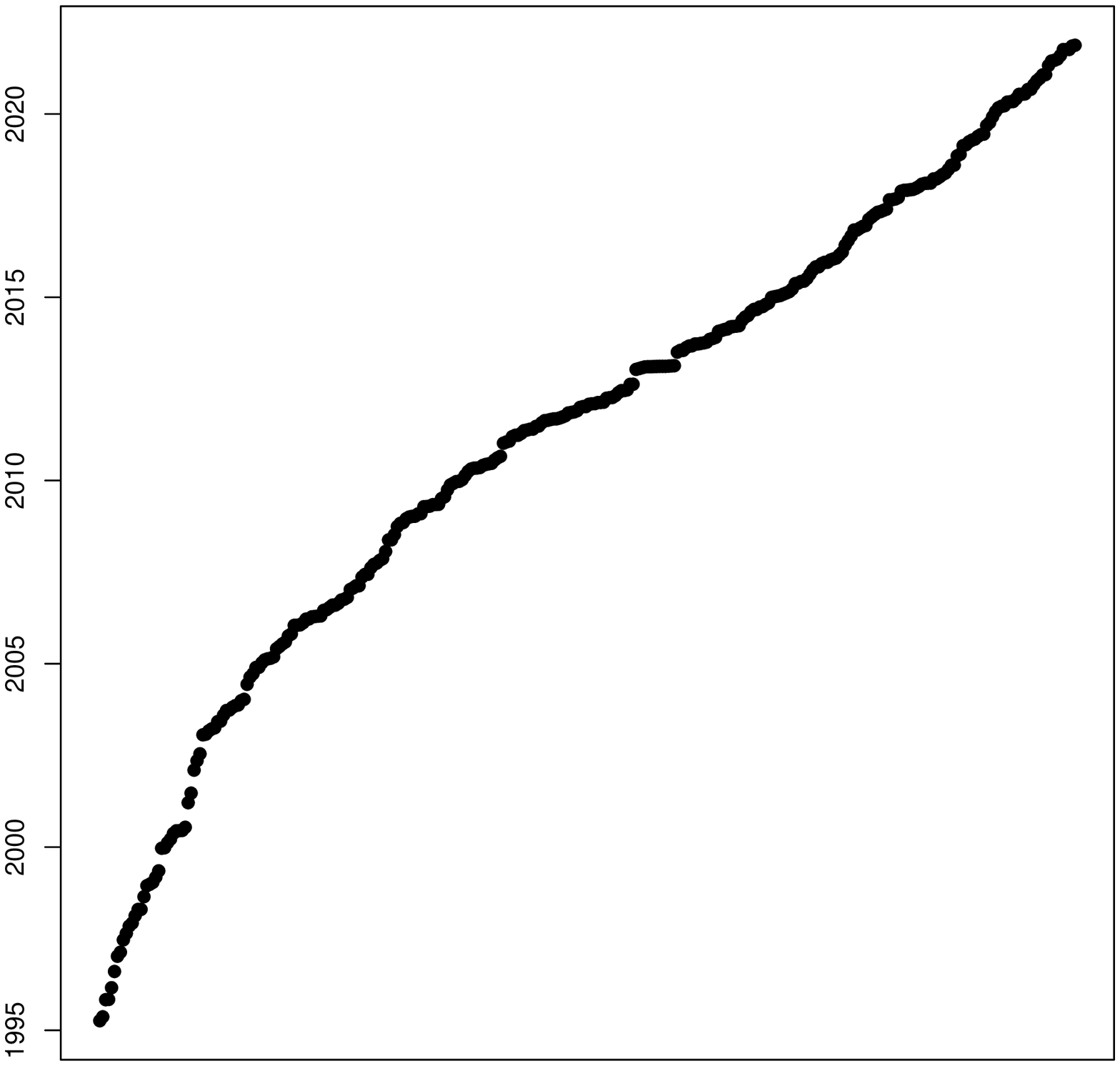}
}
\caption{Spatial (left-most panel) and temporal (right-most panel) 
projections of the $332$ earthquakes of magnitude $1.5$ or larger with 
epicentre in the Groningen gas field that occurred in the period from 
January 1st, 1995, up to December 31st, 2021.}
\label{F:eqs}
\end{figure}

\subsection{Shapefiles for the Groningen gas field}

Shapefiles for the Groningen gas field can be downloaded from the Geological 
Survey of the Netherlands TNO website  
{\tt www.nlog.nl/bestanden-interactieve-kaart}. The files are updated monthly.
In this paper we use the map that was published in April 2022.
The coordinates of the field are given in the UTM system using zone $31$ 
with metre as the spatial unit, which we rescale to kilometre. The boundary
is outlined in the left-most panel of Figure~\ref{F:eqs}.

\subsection{Earthquake catalogue}

An earthquake catalogue for The Netherlands is being maintained by the 
Royal Dutch Meteorological Office (KNMI) at \\
{\tt www.knmi.nl/kennis-en-datacentrum/dataset/aardbevingscatalogus}. \\
Data on the period before 1995 is not reliable due to the inaccuracy of 
the equipment used. Moreover, a threshold on the magnitude is necessary 
to guarantee data quality. According to Dost et al.\ \cite{Dost12}, 
for data from 1995, earthquakes with magnitude $1.5$ or larger can be reliably
recorded; a threshold of $1.3$ can be used for the period from 2010
onwards due to an extension of the monitoring network (cf.\
Hettema et al.\ \cite{Hett17}).
We use data over the time window 1995--2021 and therefore work with a
magnitude $1.5$ threshold. The coordinates of the epicentres are listed
in terms of latitude and longitude. To avoid distortions and for 
compatibility with the gas field map, we project them to UTM (zone $31$) 
coordinates. This procedure results in $332$ earthquakes, the
spatial and temporal projections of which are shown in Figure~\ref{F:eqs}.

\begin{figure}[hbt]
\centering
\centerline{ 
\epsfxsize=0.5\hsize
\epsffile{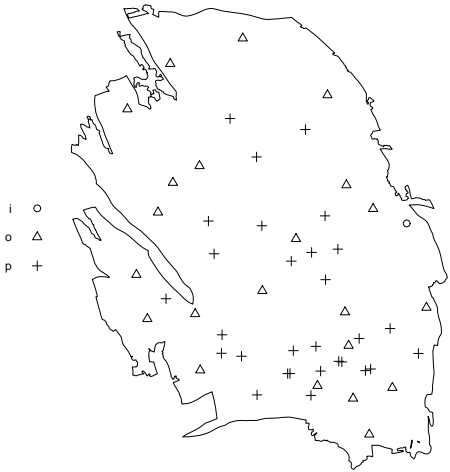}}
\caption{Production (crosses), observation (triangles) and injection (circles)
  well locations in the Groningen gas field.}
\label{F:wells}
\end{figure}

\subsection{Wells}
\label{S:location}

The exploration and production company NAM maintains a number of production,
injection and observation wells. Their coordinates (in the Amersfoort
projected coordinate system used in The Netherlands) are available from the
production plans published on \\
{\tt www.nam.nl/gas-en-olie/groningen-gasveld/winningsplan-groningen-gasveld.html}. \\
For compatibility, we transform the Amersfoort coordinates to
UTM (zone $31$).

Of the $52$ locations in the Groningen gas field shown in Figure~\ref{F:wells}, 
$29$ are production wells (indicated by a cross), one is an injection 
well (indicated by a circle) and $22$ are observation wells (indicated by a 
triangle). 

A few remarks are in order. Firstly, some production wells (at Midwolda, 
Noordbroek, Nieuw Scheemda and Uiterburen) were taken out of production 
around the year 2010 and are no longer in use. Secondly, two south-westerly 
observation wells (at Kolham and Harkstede) were drilled in a peripheral 
field rather than in the main reservoir. Finally, up to the mid 1970s, small
amounts of gas were extracted from wells not earmarked for production.


\subsection{Gas extraction}
\label{S:gas}

Monthly production values from the start of preliminary exploration
in February 1956 up to and including
December 2021 were kindly provided by Mr Rob van Eijs from Shell for all
of the $29$ production wells. The figures were given in cubic metres 
which we rescale to Nbcm. 

\begin{figure}[htb]
\centering
\centerline{
\epsfxsize=0.3\hsize
\epsffile{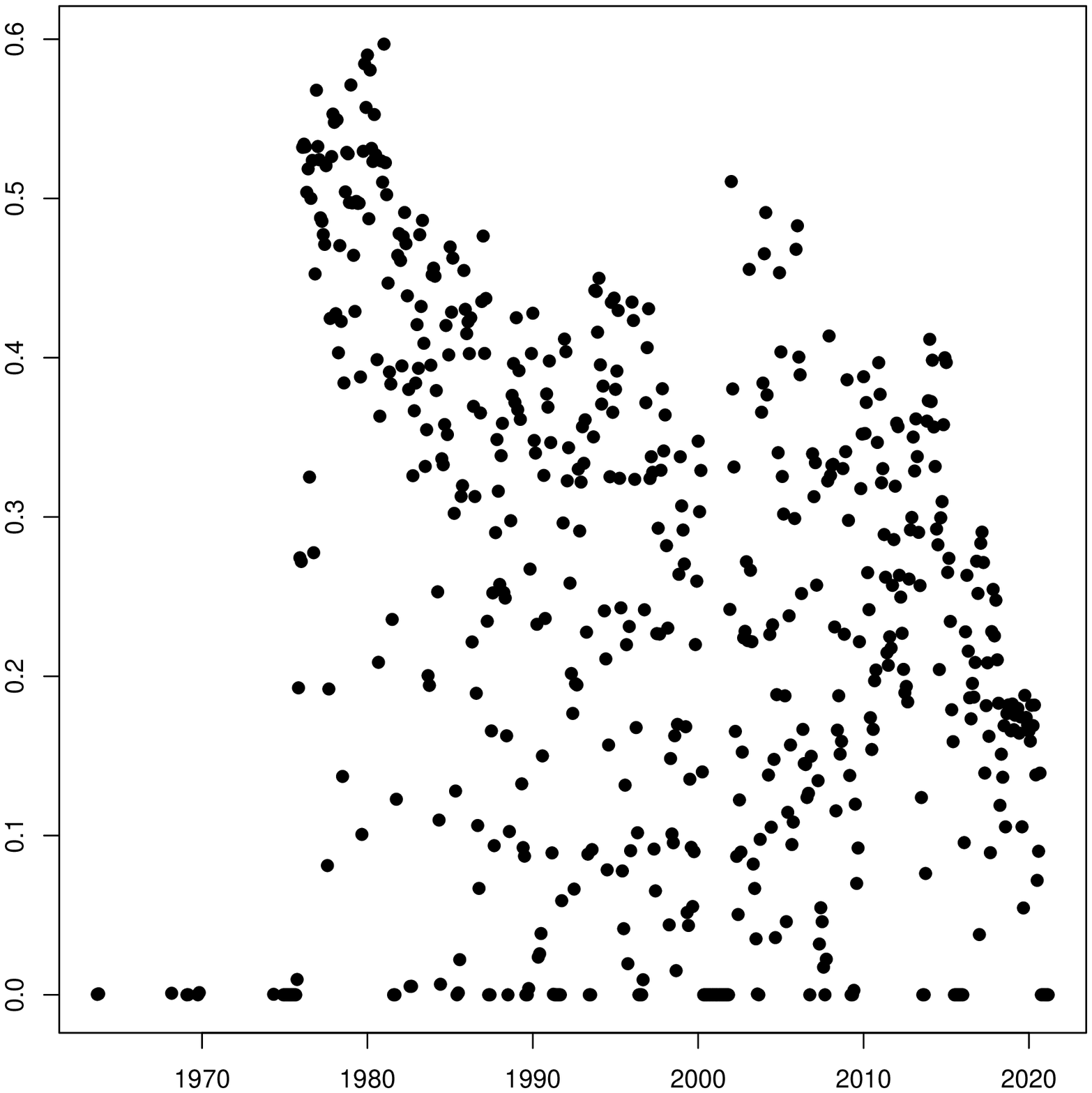}
\epsfxsize=0.3\hsize
\epsffile{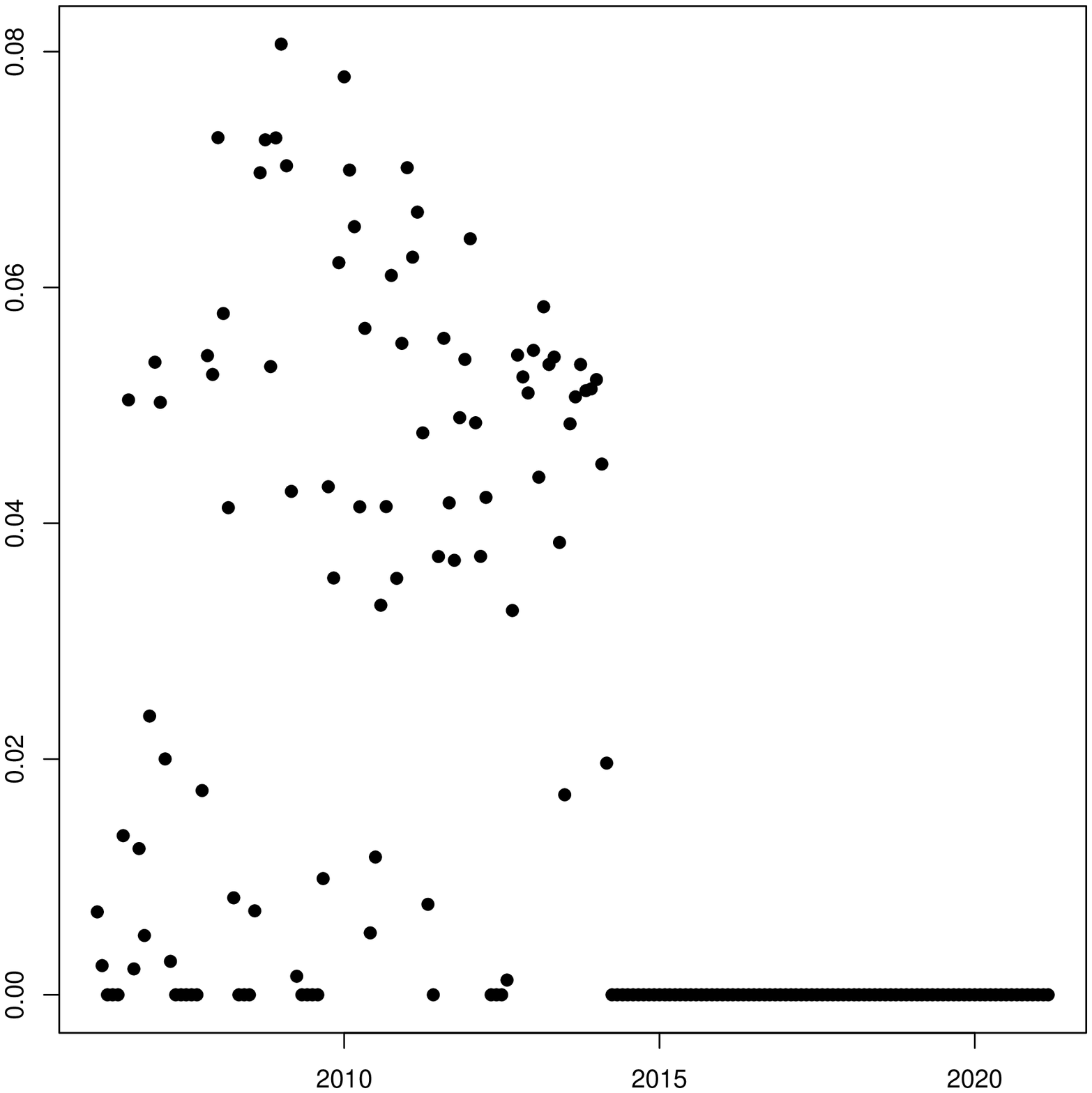}
\epsfxsize=0.3\hsize
\epsffile{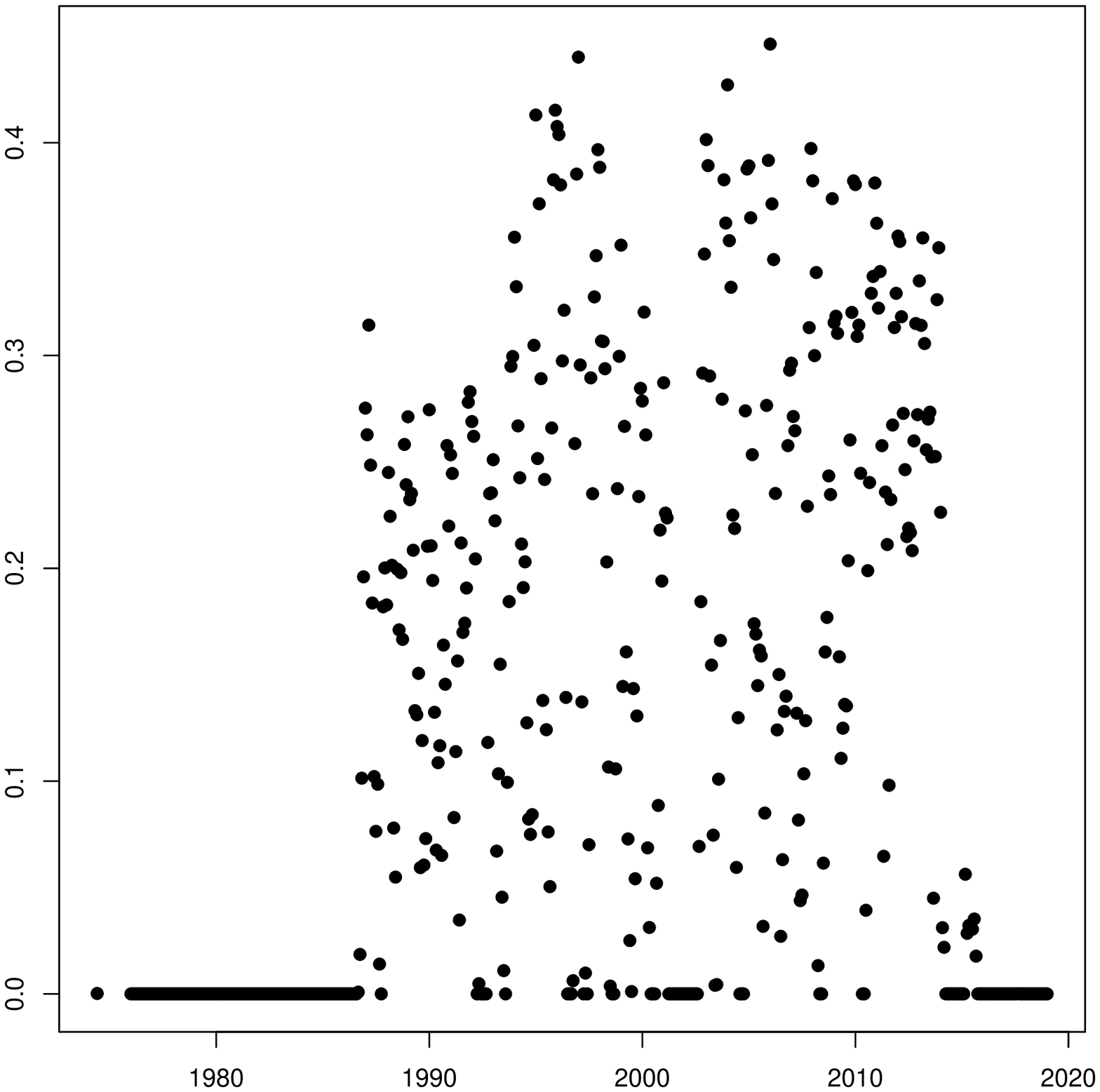}
}
\centerline{
\epsfxsize=0.3\hsize
\epsffile{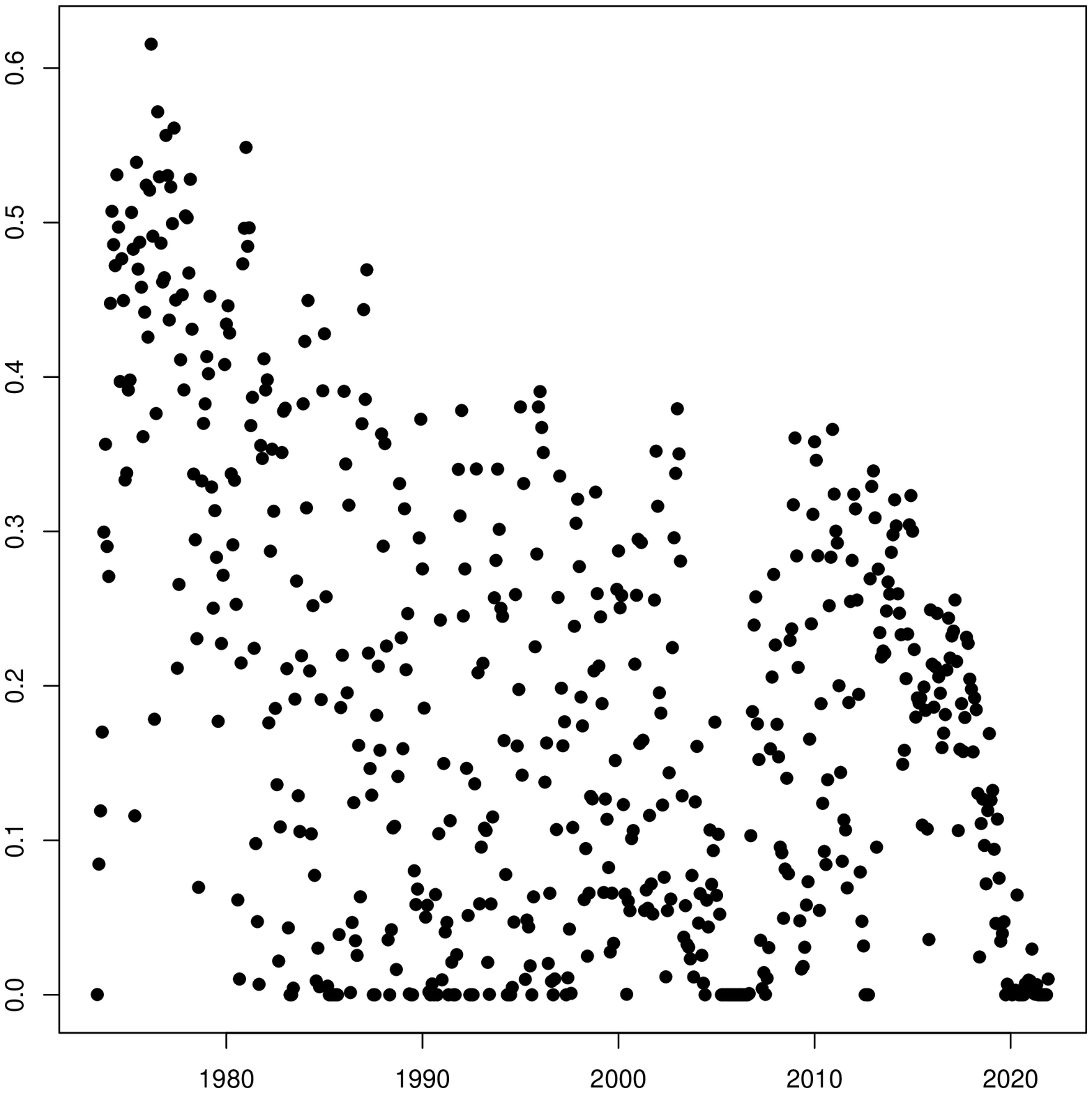}
\epsfxsize=0.3\hsize
\epsffile{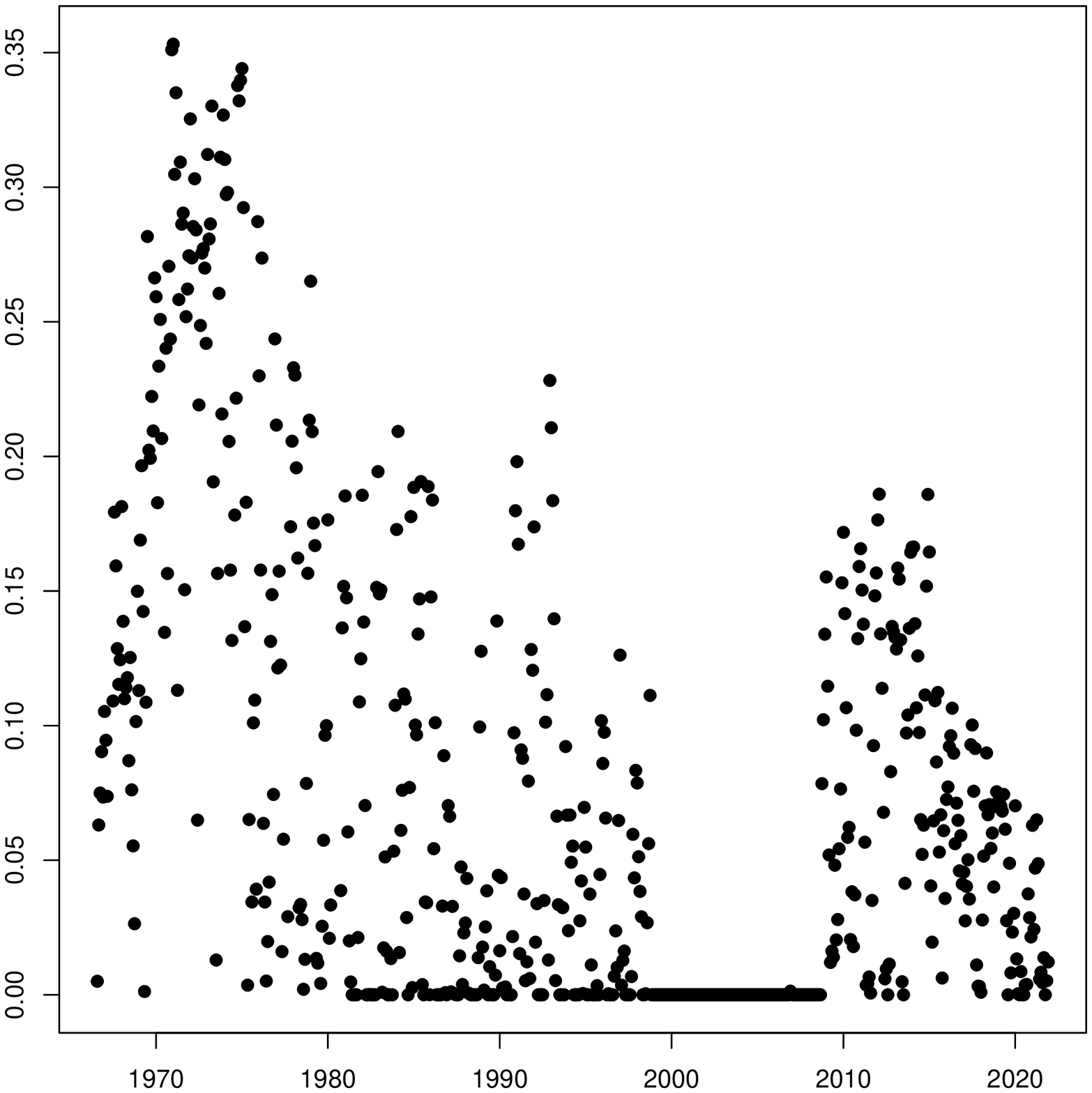}
\epsfxsize=0.3\hsize
\epsffile{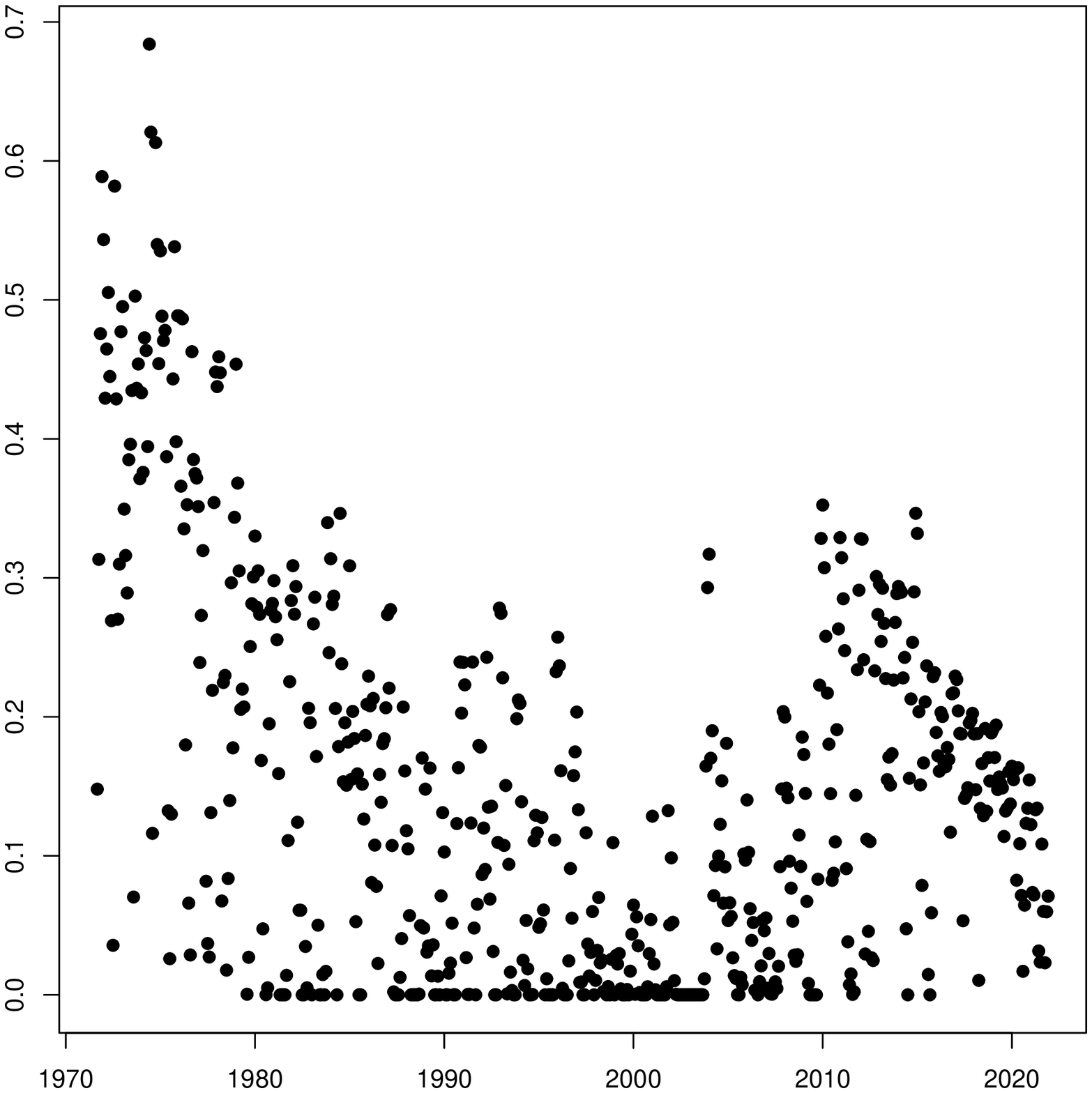}
}
\caption{Monthly production in Nbcm against time for the wells 
Bierum, Eemskanaal-13 and De Paauwen (left to right, top row)
and Amsweer, Tusschenklappen and Zuiderpolder (left to right, bottom row).}
\label{F:production}
\end{figure}

In Figure~\ref{F:production}, we show the time series for six wells chosen 
to show a range of production patterns: 
Bierum in the North-East, 
De Paauwen in the centre 
and Eemskanaal in the West 
of the gas field, Amsweer in the central East, 
Tusschenklappen in the South-West 
and Zuiderpolder in the South-East. 

One may observe that not all wells were drilled at the same time and
that some were not in use during the entire period. Also there are
differences in the amount of gas extracted: the production figures for
Eemskanaal-13 are lower than average. The sharp decline in production
from 2014 following legislation is readily apparent.

\subsection{Pore pressure observations}
\label{S:pressure}

On {\tt nam-feitenencijfers.data-app.html/gasdruk.html}, 
pore pressure observations are available over the period from April 1960 
until November 2018. In total, there are 2056 observations.
However, these data need some cleaning, as discussed in the following
paragraphs.

\begin{figure}[hbt]
\centering
\centerline{ 
\epsfxsize=0.5\hsize
\epsffile{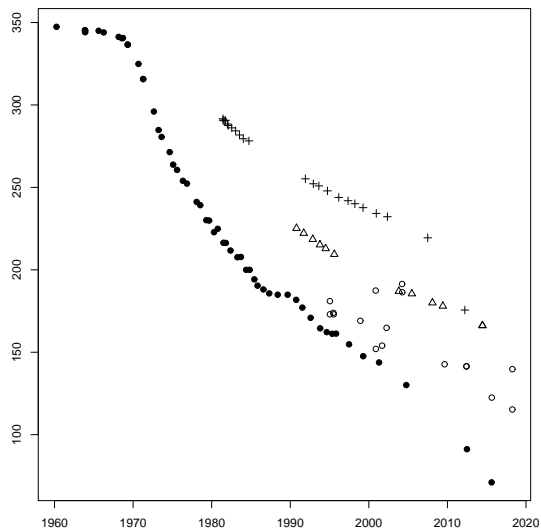}}
\caption{Pore pressure measurements in bara against time for Slochteren (discs),
Harkstede (crosses), Kolham (triangles) and Borgsweer (circles).}
\label{F:pressure}
\end{figure}

\paragraph{Errors in recorded dates}
There are some anomalies in the recorded dates. For example in
2015, the entry 10/7/2015 should be interpreted as 7/10/2015, 
July the tenth. There are eight other such errors: May 10, 2016,
May 6-7, 2017, June 5, 2017, June 7, 2017, April 9 and 11, 2018,
and November 6, 2018.

\paragraph{Missing coordinates}

Since the coordinates of one of the stations
are not listed in 
the production plans (cf.~Section~\ref{S:location}) and therefore
unknown, we omit the corresponding $28$ pore pressure 
measurements from consideration. We also disregard the two measurements 
from an observation well located outside the Groningen gas field.

The Eemskanaal-13 well is the only one depleting a peripheral field,
the so-called Hark\-stede block. Moreover, as can be seen from 
Figure~\ref{F:production}, it is extracting less gas than other wells. 
The combined effect is that the pore pressure measurements are somewhat higher 
than at other wells. According to an expert, setting its location to 
either that of the Eemskanaal plant or to that of the installation at
Harkstede would lead to biases and it is therefore preferable to ignore
the seven observations for Eemskanaal-13 altogether.

\paragraph{Invalid measurements}

The NAM file mentions ten cases in which the observations are invalid
for various reasons. We delete these measurements.


After cleaning, we are left with $2009$ pore pressure measurements.
Figure~\ref{F:pressure} shows time series at four locations: 
a production well in the South-West (Sloch\-te\-ren), two
observation wells (Harkstede and Kolham) in peripheral fields and 
the injection well at Borgweer in the East. Mostly, the graph is initially
flat, followed by a decrease. Note that the measurements
in the periphery are higher than those in the main reservoir. 


\section{Exploratory data analysis}
\label{S:explore}

In this paper, we treat the earthquake catalogue as a 
spatio-temporal point pattern, a realisation of a point process
in space and time. Formally, let $\Psi$ be a simple spatio-temporal 
point process in $W_S\times W_T$ for bounded open sets $W_S \subset
\oR^d$ and $W_T\subset \oR$ and suppose that its first order moment
measure exists, is finite and absolutely continuous with respect to
the product of Lebesgue measures $\ell$ in space and time 
(see e.g.\ Chiu et al.\ \cite{SKM}).
Write $\lambda$ for its Radon--Nikodym derivative, known as
the {\em intensity function\/},  and $N(\Psi \cap
(W_S\times W_T))$ for the number of points placed by $\Psi$ in
$W_S \times W_T$. Intuitively speaking, $\lambda(s,t) ds dt$ is the
probability that $\Psi$ places a point in the infinitesimal region
$ds dt$ around $(s,t) \in W_S \times W_T$.

As usual in spatial statistics, we start by an empirical exploration
of the trend and interaction.

\subsection{Adaptive kernel estimation of the intensity function}
\label{S:intensity}

Our first step is to estimate the intensity function of the point 
process of earthquakes. The standard technique to do so is kernel
estimation as proposed by Diggle \cite{Digg85}. This technique can be 
seen as a generalisation
of the histogram. Briefly, a Gaussian kernel (say) with a given
standard deviation $h_S$ in space and $h_T$ in time is centred at each
of the points in a realisation of $\Psi$ and their sum is reported.
The choice of the {\em bandwidhts\/} $h_S$ and $h_T$ is crucial.
Some asymptotical results are available in Chac\'on and Duong 
\cite{ChacDuon18}, Van Lieshout \cite{Lies20} and Lo \cite{Lo17},
but practical rules of thumb seem to be lacking.

\begin{figure}[hbt]
\centering
\centerline{
\epsfxsize=0.45\hsize
\epsffile{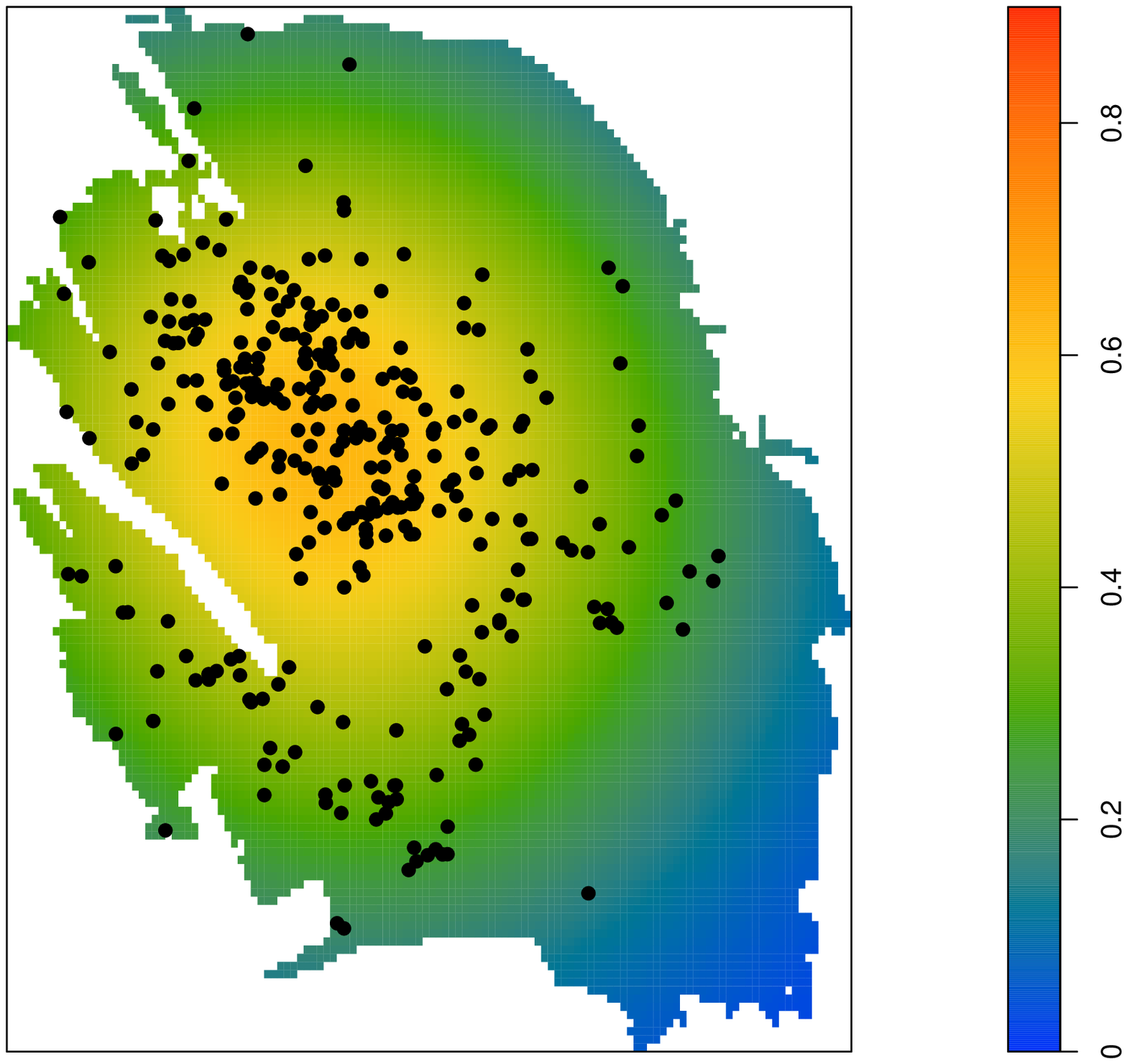}
\epsfxsize=0.45\hsize
\epsffile{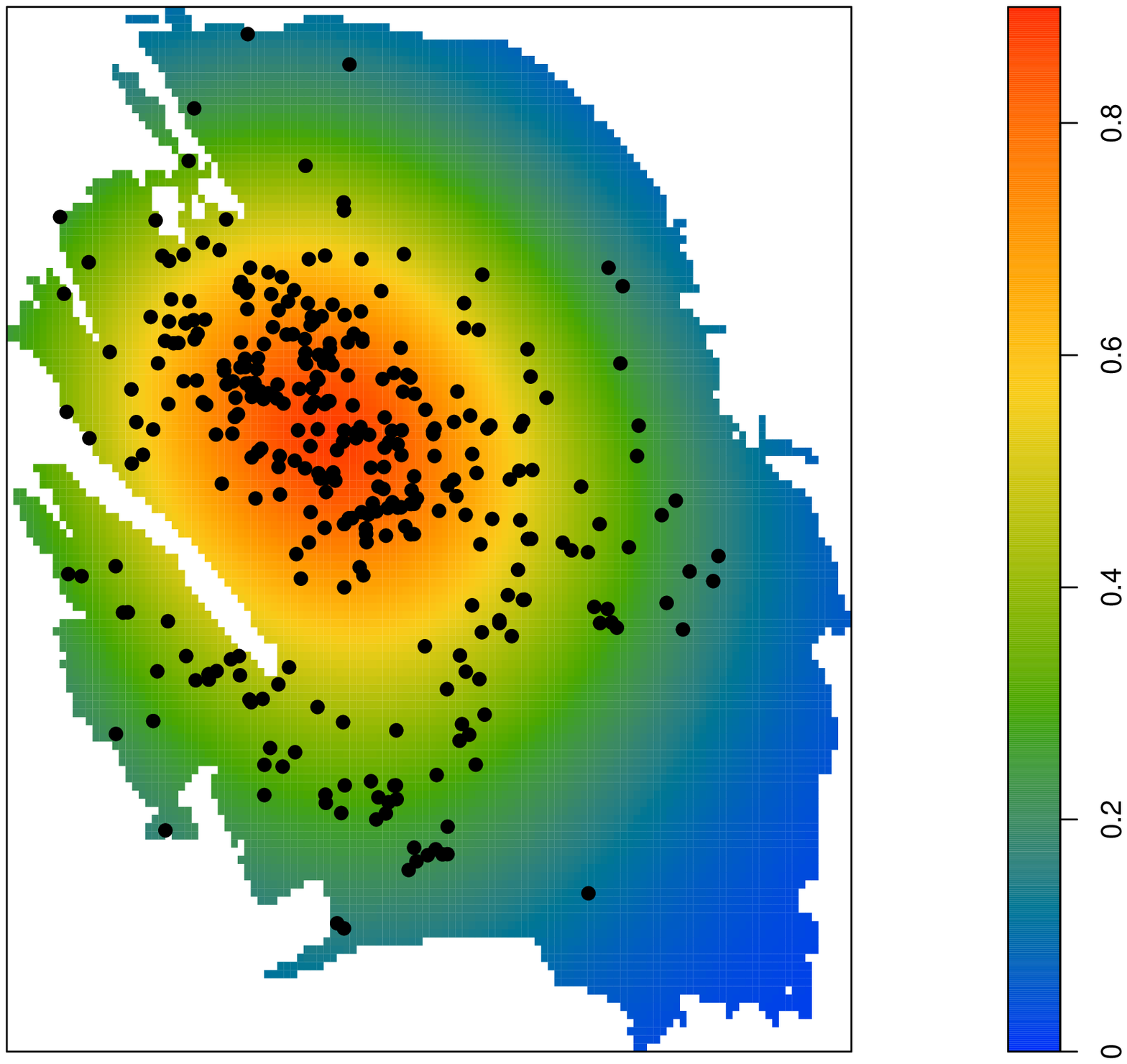}}
\caption{Projections on space of classic (left-most panel) and 
adaptive (right-most panel) kernel estimates of intensity (per square 
km) for the data in Figure~\ref{F:eqs}. The pilot bandwidths are $h_{g,S} 
= 9.4$ and $h_{g,T} = 182.5$, whilst $h_{a,S} = 6.9$ and $h_{a,T} = 212.9$}
\label{F:intensitySpace}
\end{figure}

The risk of using the same bandwidths $h_S, h_T$ at all points
of $\Psi$ is that, as all one-size-fits-all solutions, this approach
tends to oversmooth in regions that are rich in points while at the
same time it does not smooth enough in sparser regions. To overcome
this drawback we propose to use an adaptive smoother as introduced for
classic random variables by Abramson \cite{Abra82a}. In a spatial 
context, such estimators were studied by Davies et al.\ \cite{Davi18} 
for Poisson point processes and in Van Lieshout \cite{Lies21} for general
point processes with interaction between the points. The underlying
idea is to weigh the bandwidth at $(s,t) \in \Psi$ by a scalar that is
inversely proportional to $\sqrt{ \lambda(s,t)}$. The power $1/2$ is
motivated by asymptotics (cf.\ Abramson \cite{Abra82a} and Van Lieshout
\cite{Lies21}), in practice other powers could also be used.


Formally, set
\begin{equation}
\label{e:lambdaA}
\hat \lambda_A(x_0; h_S, h_T) =
\sum_{(s,t)\in \Psi \cap (W_S\times W_T)} 
\frac{
\kappa( H(c(s,t) h_S, c(s,t) h_T)^{-1} (x_0-(s,t)) )}{
w( ((s,t), h_S, h_T ) \, c(s,t)^{3} \, h_S^2 \, h_T}
\end{equation}
for $x_0 \in W_S\times W_T$. Here $H(h_1, h_2)$, $h_1, h_2 > 0$, 
is a $3\times 3$ diagonal matrix whose first two entries are $h_1$ and 
whose third entry is $h_2$,
\begin{equation}
\label{e:geomMean}
c(s,t) = \left(
\frac{\lambda(s,t)}{
( \prod_{z\in \Psi \cap (W_S\times W_T)} 
  \lambda(z) )^{1/N(\Psi\cap W_S\times W_T)}
} 
\right)^{-1/2},
\end{equation}
$w( (s,t), h_S, h_T)$ is an edge correction term
and $\kappa$ is a symmetric probability density function, 
the kernel.
Since $\lambda$ is unknown, $c(s,t)$  must be estimated. We will do
so by plugging in a standard kernel estimator with fixed bandwidth.

To select the bandwidth, we propose the following completely
non-parametric and computationally easy algorithm.

\begin{algo}  \label{A:Hamilton}
Let $\psi$ be a non-empty spatio-temporal point pattern that is
observed in the window $W_S\times W_T$. Then
\begin{enumerate}
\item Choose global bandwidth $h_{g,S}$, $h_{g,T}$ 
by minimising
\begin{equation} 
\left| \sum_{x \in \psi \cap (W_S\times W_T)} 
\frac{1}{ \hat \lambda(x; h_S, h_T) } - \ell(W_S) \ell(W_T)
 \right|
\label{e:classicHamilton}
\end{equation}
over $h_S, h_T >0$ (with minimal $h_S^2 h_T$ in case of multiple
solutions) where
\[
\hat \lambda(x; h_S, h_T) = 
\frac{1}{h_S^2 h_T} \sum_{y \in \psi \cap (W_S\times W_T)} 
     \kappa\left( H(h_S, h_T)^{-1} ( x - y ) \right).
\]
\item Calculate, for each $x\in\psi \cap (W_S\times W_T)$, the edge-corrected 
pilot estimator 
\[
\hat \lambda_c(x; h_{g,S}, h_{g,T} ) = 
\frac{1}{h_{g,S}^2 h_{g,T}} \sum_{y \in \psi \cap (W_S\times W_T)} 
\frac{
     \kappa\left( H(h_{g,S},h_{g,T})^{-1} ( x - y ) \right)
}{
w(y, h_{g,S}, h_{g,T})
}
\]
with local edge correction weights
\[
w(y, h_{g,S}, h_{g,T}) = 
\frac{1}{h_{g,S}^2 h_{g,T}} \int_{W_S\times W_T} 
\kappa\left( H(h_{g,S},h_{g,T})^{-1} (z - y ) \right) dz.
\]
\item Choose adaptive bandwidths 
$h_{a,S}$, $h_{a,T}$, by minimising
\[
\left| \sum_{x \in \psi \cap (W_S\times W_T)} 
\frac{1}{ \hat \lambda_A(x; h_S, h_T) }
  - \ell(W_S) \, \ell(W_T) \right|
\]
over $h_S, h_T >0$ (with minimal $h_S^2 h_T$ in case of multiple
solutions) where $\hat \lambda_A$ is given by (\ref{e:lambdaA}) 
and (\ref{e:geomMean}) with $w \equiv 1$ upon plugging in the 
edge-corrected pilot estimator $\hat \lambda_c$ for $\lambda$.
\end{enumerate}
\end{algo}

Local edge correction weights, as suggested by Van Lieshout \cite{Lies12}, 
for the adaptive kernel estimator take the form
\[
w(y, h_{a,S}, h_{a,T}) = 
\int_{W_S\times W_T}
\frac{
 \kappa\left( H( \hat c(y) \, 
                   (h_{a,S}, h_{a,T}) )^{-1} (z-y)  \right) 
}{
h_{a,S}^2 h_{a,T} (\hat c(y))^{3} } \, dz
\]
and ensure that (\ref{e:lambdaA}) is mass preserving in the sense
that its integral over $W_S\times W_T$ is equal to the number of points
in $\psi\cap (W_S\times W_T)$. In selecting the
bandwidth in steps 1 and 3 of Algorithm~\ref{A:Hamilton}, no edge
correction is applied in order to obtain a clear optimum (see
Cronie and Van Lieshout \cite{CronLies18}). 

The justification of Algorithm~\ref{A:Hamilton} lies in the
Campbell--Mecke theorem (cf.\ Chiu et al.\ \cite{SKM}), which states that
\[
\oE \left[ \sum_{x \in \Psi \cap (W_S\times W_T)} \frac{1}{\lambda(x)}
  \right] = \ell(W_S) \ell(W_T).
\]
The first step is a modification to space-time of the 
Cronie--Van Lieshout bandwidth selector \cite{CronLies18} for purely
spatial point processes. An important difference is that in space one
optimises over a single parameter and usually, but not always, there
is only one minimiser. In the space-time domain, the optimisation is
done with respect to two parameters, $h_S$ and $h_T$. Therefore, as a
rule, a curve of minimisers is found in steps 1 and 3 of
Algorithm~\ref{A:Hamilton}.
We pick the optimiser having the smallest scale $h_S^2 h_T$.

\begin{figure}[hbt]
\centering
\centerline{
\epsfxsize=0.45\hsize
\epsffile{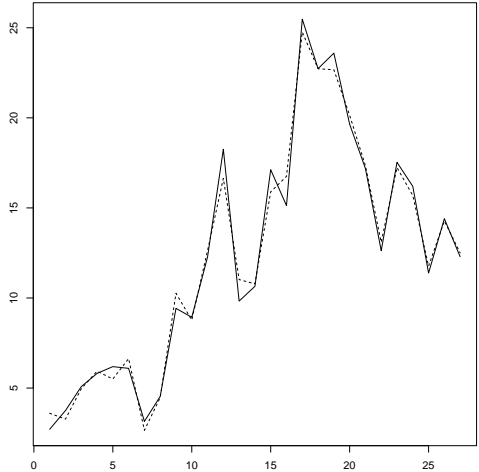}}
\caption{Expected number of earthquakes using classic (broken line) 
and adaptive (solid line) kernel estimates for the data in 
Figure~\ref{F:eqs} over $27$ equal time intervals. The pilot 
bandwidths are $h_{g,S} = 9.4$ and $h_{g,T} = 182.5$, whilst 
$h_{a,S} = 6.9$ and $h_{a,T} = 212.9$}
\label{F:intensityTime}
\end{figure}

For our earthquake catalogue, the results are shown in
Figures~\ref{F:intensitySpace} and \ref{F:intensityTime}.
The pilot bandwidths are $h_{g,S} = 9.4$ and $h_{g,T} = 182.5$.
The corresponding edge-corrected projections on space and time 
are shown in, respectively, the left-most panel of 
Figure~\ref{F:intensitySpace} and the broken line in 
Figure~\ref{F:intensityTime}. They should be compared to the projections
of the edge-corrected adaptive kernel estimate with $h_{a,S} = 6.9$ 
and $h_{a,T} = 212.9$ shown in the right-most panel in 
Figure~\ref{F:intensitySpace} and the solid line in 
Figure~\ref{F:intensityTime}. Note that $\hat\lambda_A$ attains
higher values in the central reservoir than $\hat\lambda$, lower
values near the eastern border of the gas field. From a temporal
perspective, the extremes in years with a large number of earthquakes
are somewhat more pronounced for the adaptive kernel estimate.

\subsection{Inhomogeneous spatio-temporal K- and J-functions}

Having estimated  the trend, we now turn our attention to
the inter-point interactions. To quantify such interaction,
information about the joint distributions of pairs, or tuples,
of points is required, which is formalised by the higher-order
analogues of the intensity function, the
product densities $\lambda^{(n)}$. Heuristically,
$\lambda^{(n)}( (s_1, t_1), \ldots, (s_n, t_n) ) ds_1 dt_1
\cdots ds_n dt_n$ is the probability that $\Psi$ places points
at each of the infinitesimal regions $ds_i dt_i$ around
$(s_i, t_i)$, $i=1, \dots, n$.

A spatio-temporal point process $\Psi$ on $\oR^2 \times \oR$ is said 
to be intensity-reweighted moment stationary (IRMS) (cf.\ Cronie and 
Van Lieshout \cite{CronLies15}) if its product densities $\lambda^{(n)}$ 
of all orders exist, 
$\bar{\lambda} = \inf_{(s,t)}\lambda(s,t)>0$ and, for all $n\geq1$, 
$\xi_n$ is translation invariant in the sense that 
$$
\xi_n((s_1,t_1) + (a,b), \ldots, (s_n,t_n) + (a,b)) = 
\xi_n((s_1,t_1), \ldots, (s_n,t_n)) 
$$
for almost all $(s_1,t_1), \ldots, (s_n,t_n) \in \oR^{2}\times\oR$ and all 
$(a,b) \in \oR^{2}\times\oR$. Here $\xi_n$ are the 
$n$-point correlation functions defined 
in terms of the $\lambda^{(n)}$ by setting $\xi_1 \equiv 1$ and for other
$n$ recursively by
\begin{align}
\label{CorrFunc}
\frac{
\lambda^{(n)}((s_1,t_1), \ldots, (s_n,t_n))
}
{\prod_{k=1}^{n} \lambda(s_k,t_k)}
&=
\sum_{k=1}^{n}
\sum_{D_1, \ldots, D_k}
\prod_{j=1}^{k}
\xi_{|D_j|}( (s_i,t_i): i \in D_j ),
\end{align}
where 
$\sum_{D_1, \ldots, D_k}$ is a sum over all possible $k$-sized partitions 
$\{D_1, \ldots, D_k \}$, $D_j \neq \emptyset$, of the set $\{ 1, \ldots, n \}$ 
and $|D_j|$ denotes the cardinality of $D_j$. For a Poisson point process, 
in which there are no correlations between the points, $\xi_n \equiv 0$ 
for $n\geq 1$.  The point process is said to be second order 
intensity-reweighted stationary (SOIRS) if the translation invariance 
holds up to $n=2$ (see Gabriel and Diggle \cite{GabrDigg09}).

Various summary statistics exist to explore inter-point interactions.
Write
\[
S_{h_S}^{h_T} = \{ (s, t) \in \oR^2\times\oR: \| s \| \leq h_S, |t| \leq h_T \}
\]
and let $\Psi$ be an IRMS spatio-temporal point process. Then the 
statistic
\[
J_n(h_S, h_T) 
= \int_{S_{h_S}^{h_T}} \cdots \int_{S_{h_S}^{h_T}} 
\xi_{n+1}((0,0), (s_1,t_1), \ldots, (s_n,t_n)) 
\prod_{i=1}^{n}ds_i dt_i 
\]
quantifies cumulative correlations of order $n =1, 2, \dots$
up to ranges $h_S \geq 0$ in space and $h_T \geq 0$ in time. 
Multiple orders can be combined. For example, Van Lieshout \cite{Lies11} proposed
\begin{align}
\label{JfunSTPP}
J_{\rm inhom}(h_S, h_T) = 1 + \sum_{n=1}^{\infty} \frac{(-\bar{\lambda})^n}{n!} 
   J_n(h_S, h_T) 
\end{align}
for all spatial ranges $h_S \geq 0$ and temporal ranges $h_T \geq 0$ for which 
the series is absolutely convergent. Values greater than one are indicative
of repulsion, values smaller than one suggest clustering. For further
details, the reader is referred to Cronie and Van Lieshout \cite{CronLies15}. Truncating at $n=2$,
\[
J_{\rm inhom}(h_S, h_T) - 1
\approx - \bar{\lambda} \left( K_{\rm inhom}(h_S, h_T) - \ell(S_{h_S}^{h_T}) 
\right)
\]
in terms of the inhomogeneous $K$-function 
\[
 K_{\rm inhom}(h_S, h_T) = \int_{S_{h_S}^{h_T}} 
\left[ 
\xi_2( (0,0), (s_1, t_1)) + 1
\right] ds_1 dt_1
\]
of Gabriel and Diggle \cite{GabrDigg09}, which is well-defined under the 
SOIRS assumption. Values greater or smaller than the volume of 
$S_{h_S}^{h_T}$ are indicative of, respectively, clustering and 
inhibition between points.

\begin{figure}[hbt]
\centering
\centerline{
\epsfxsize=0.45\hsize
\epsffile{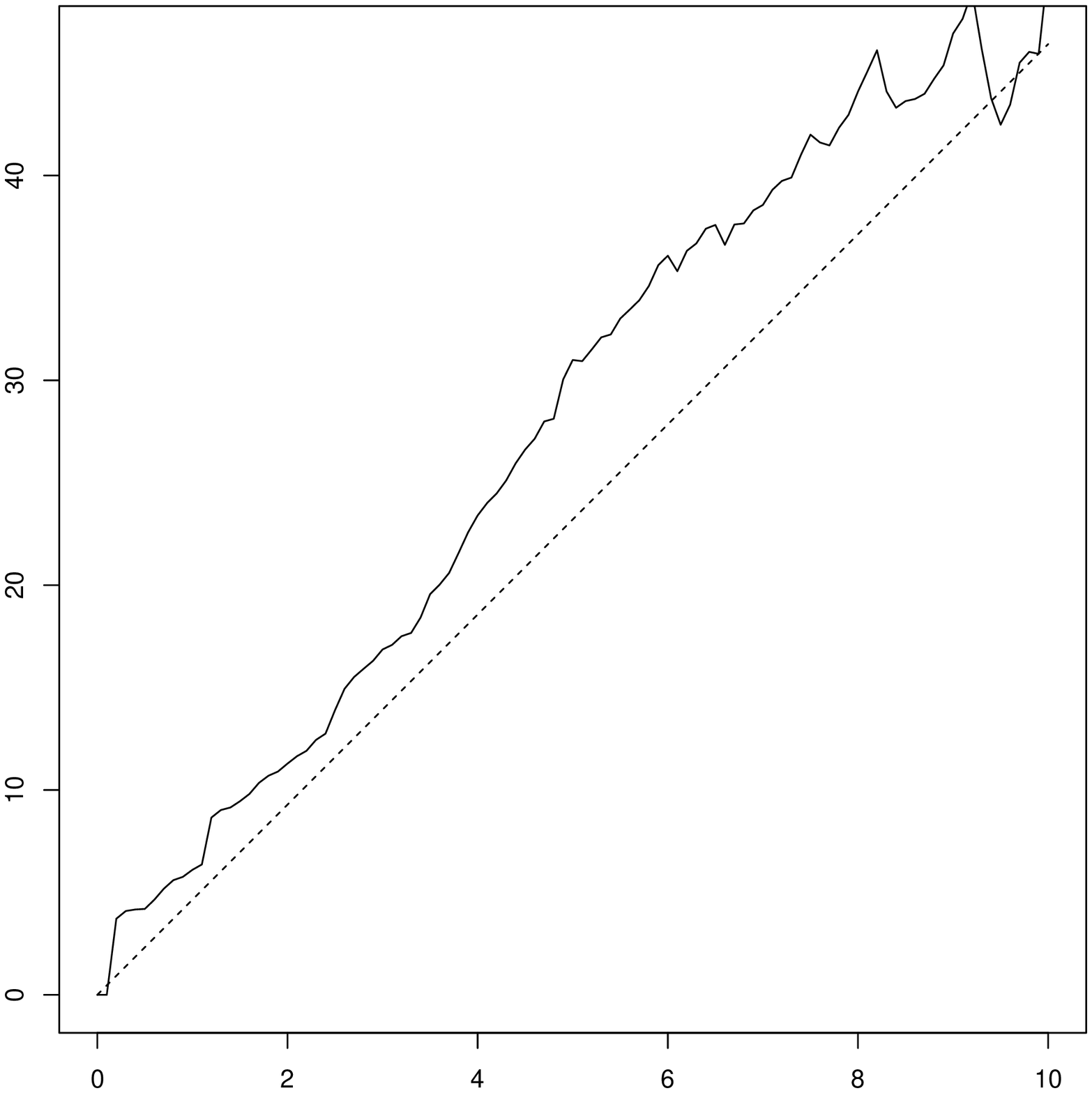}
\epsfxsize=0.45\hsize
\epsffile{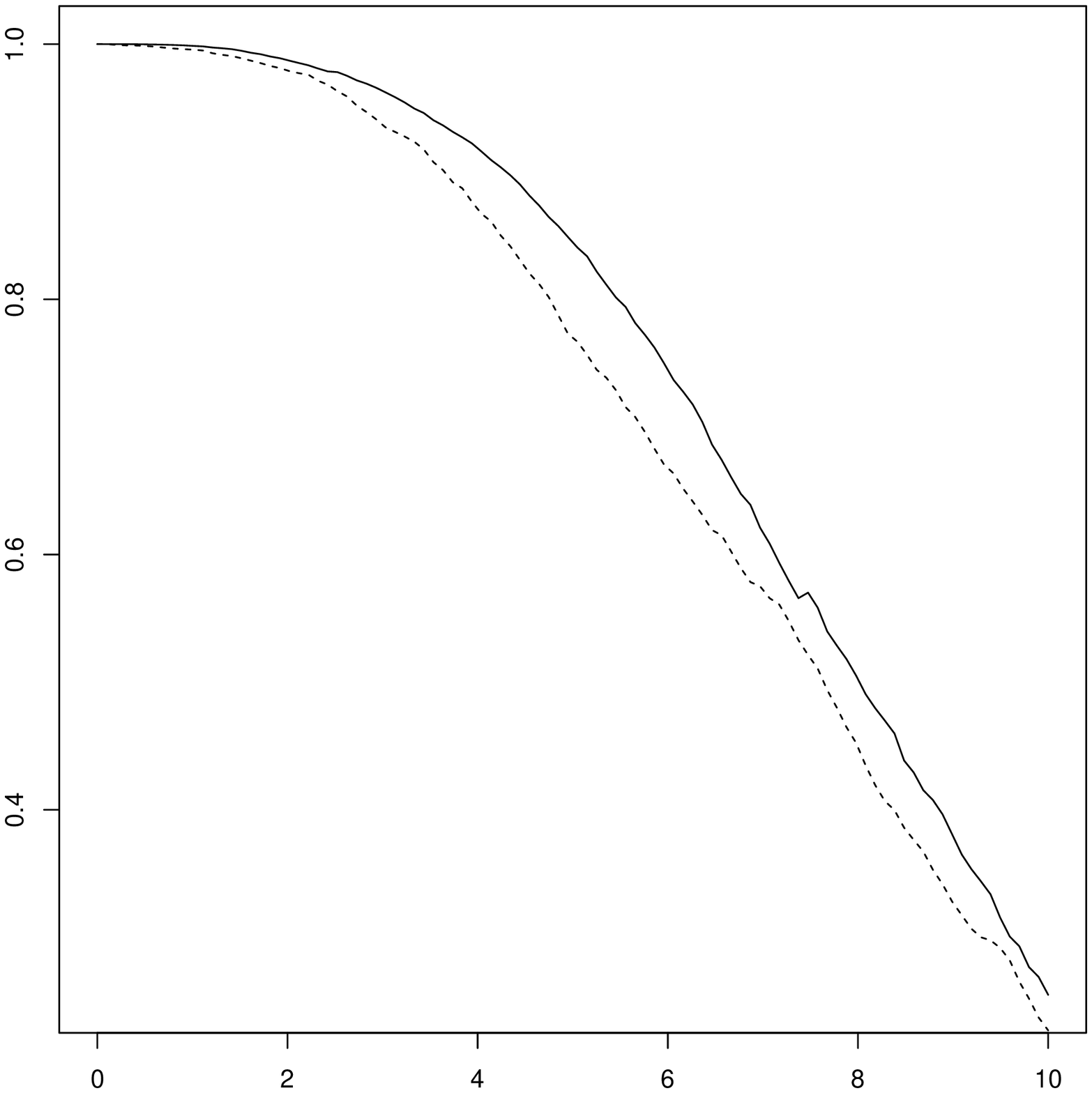}}
\caption{Graphs of $( \hat K_{\rm{inhom}}(r, 100 r) / (2\pi) )^{1/3} $
against $r$ (solid line, left-most panel), of 
$\hat G(1-u^y_{r, 100r})$ (solid line, right-most panel) and 
$G^{!y}(1-u^y_{r, 100r})$ (broken line, right-most 
panel) for the data in Figure~\ref{F:eqs}.
The units are km in space and days in time.}
\label{F:FGK}
\end{figure}

In practice, one must estimate these statistics based on a
pattern observed in $W_S\times W_T$. The definition of the 
inhomogeneous $J$-function does not immediately suggest a 
suitable estimator. However, Cronie and Van Lieshout \cite{CronLies15}
showed that, for all $y \in \oR^2 \times \oR$,
\begin{equation} \label{e:Jgenfun}
J_{\rm inhom}(h_S, h_T)
=
\frac{G^{!y}(1-u_{h_S,h_T}^{y})}{G(1-u_{h_S,h_T}^{y})}
\end{equation}
whenever well-defined and the denominator is greater than zero. 
Here the function $u_{h_S, h_T}^{y}$ is defined by
\[
u_{h_S,h_T}^{y}(s,t)
=
\frac
{
\bar{\lambda}
1\{ \| a - s \| \leq h_S, | b - t |\leq h_T \}
}
{\lambda(s,t)},
\quad y = (a,b) \in\oR^2\times\oR,
\]
and $G$ is the generating functional of $\Psi$, that is,
\[
 G(1-u_{h_S,h_T}^{y})
 =   \oE\left[
\prod_{(s,t) \in \Psi} \left(
1 -
\frac{\bar{\lambda} 1\{ \| a - s \| \leq h_S, | b - t | \leq h_T \}
}{\lambda(s,t)}
\right)
\right]
\]
for $h_S, h_T \geq 0$, 
under the convention that empty products take the value one. 
$G^{!y}$ is defined similarly in terms of the distribution of
$\Psi \setminus \{y\}$ given there is a point of $\Psi$ at $y$.
Being expectations, the numerator and denominator in equation
(\ref{e:Jgenfun}) can be estimated in a straightforward manner.
Write $W_S^{\ominus {h_S}}$ for the set of points in $W_S$
that are at least $h_S$ away from the border of $W_S$ and
let $W_T^{\ominus h_T}$ be the similarly eroded temporal domain.
Then, 
given a finite point grid $L \subseteq W_S \times W_T$,
\[
\frac{1}{ N(L \cap (W_S^{\ominus h_S} \times W_T^{\ominus h_T} ) ) }
\sum_{l\in L \cap (W_S^{\ominus h_S} \times W_T^{\ominus h_T})}
\left[
\prod_{x \in \Psi \cap ( l + S_{h_S}^{h_T} )}
\left(1 - \frac{\bar{\lambda}}{\lambda(x)}\right)
\right]
\]
is an unbiased estimator of the denominator in (\ref{e:Jgenfun}).
An unbiased estimator for the numerator is obtained analogously.
Finally,
\[
\hat{K}_{\rm{inhom}}(h_S,h_T)
 =  \frac{1}{ \ell( W_S^{\ominus h_S}) \ell(W_T^{\ominus h_T}) }
\sum_{x \in \Psi \cap (W_S^{\ominus h_S} \times W_T^{\ominus h_T})}
 \sJM_{y \in \Psi \cap (x + S_{h_S}^{h_T}) } \frac{1}
  {\lambda(x)\lambda(y) }
\]
is an unbiased estimator of $K_{\rm{inhom}}(h_S, h_T)$. 
Note that the intensity functions are unknown. A practical solution is
to plug in their estimated counterparts (cf.\ Section~\ref{S:intensity}).

For the earthquake catalogue depicted in Figure~\ref{F:eqs}, 
consider Figure~\ref{F:FGK}. The solid line in the left-most panel
of the figure is the graph of 
$( \hat K_{\rm{inhom}}(r, 100 r) / (2\pi) )^{1/3}$ for the
earthquake data. It lies above the graph of the same function for a
Poisson point process (shown as a broken line in the 
left-most panel of the figure), suggesting attraction between the 
points. In the right-most panel of Figure~\ref{F:FGK}, the graph of
$\hat G(1-u^y_{r, 100r})$ estimated from the data lies above that
of $G^{!y}(1-u^y_{r, 100r})$, which confirms the suggested clustering.

\section{Explanatory variables} 
\label{S:covariates}

Gas production and pore pressure (cf.\ Section~\ref{S:gas} and
\ref{S:pressure}) may well have an effect on the earthquake rates.
However, they are measured only for a limited number of locations
and times. Therefore, in this session, we discuss how to
calculate appropriate values  for the entire space-time domain.

\subsection{Non-parametric smoothing of gas production values}

Monthly production values are available for the $29$ production 
wells shown by a cross in Figure~\ref{F:wells} over the time period 
1956--2021. The gas extracted from the Eemskanaal-13 bore hole 
is listed separately from the other ones at the Eemskanaal site.
The reason is that the Eemskanaal-13 pipe is not drilled vertically
but bends underground in such a way that it is depleting the
Harkstede block (cf.\ Section~\ref{S:pressure}). Therefore we assign its 
production to the location of the observation well at Harkstede. 
Thus, we end up with a spatial pattern of $30$ points, the locations 
of the $29$ production wells as well as the Harkstede proxy for the 
Eemskanaal-13 pipe.

\begin{figure}[htb]
\centering
\centerline{
\epsfxsize=0.45\hsize
\epsffile{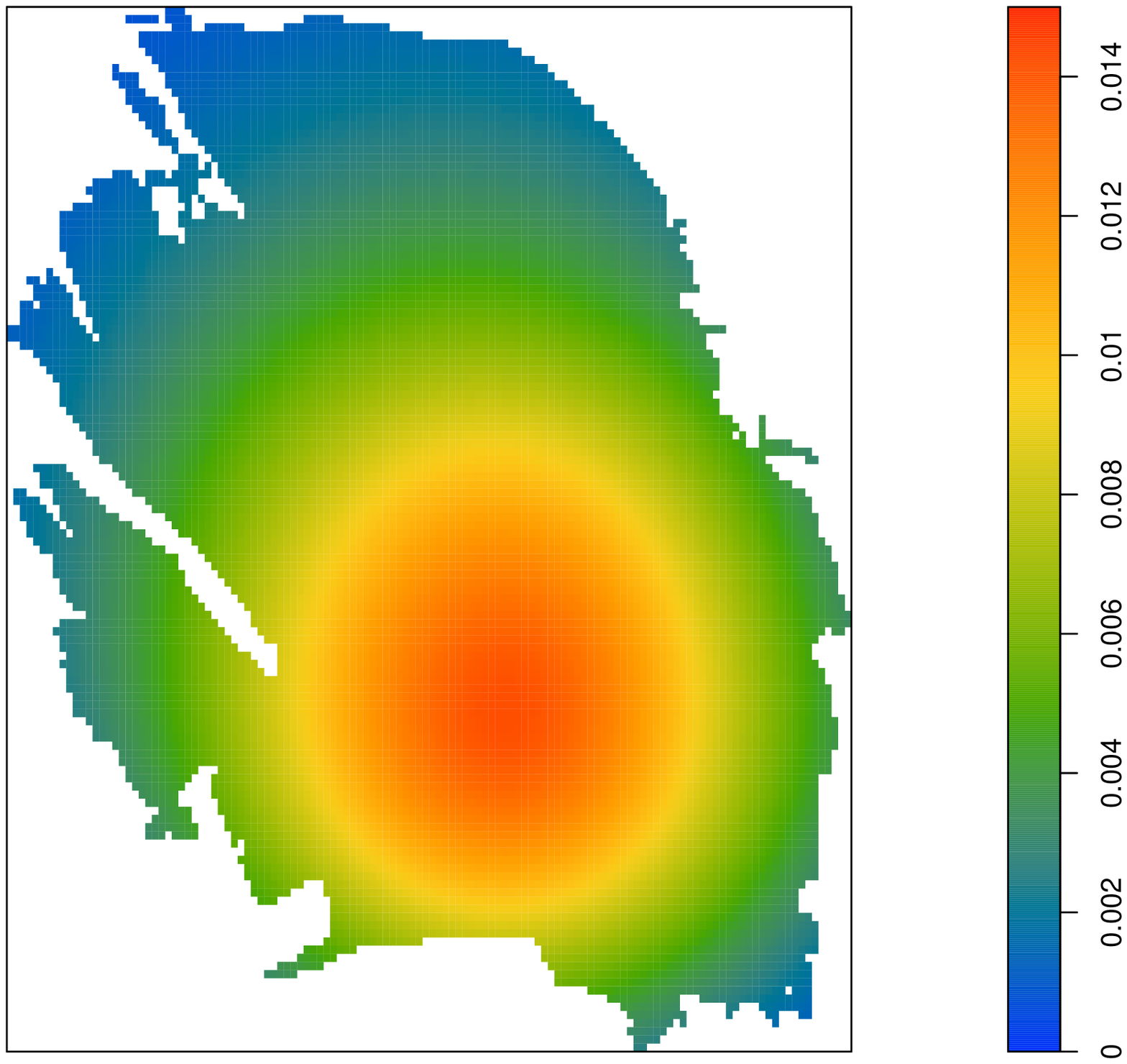}
\epsfxsize=0.45\hsize
\epsffile{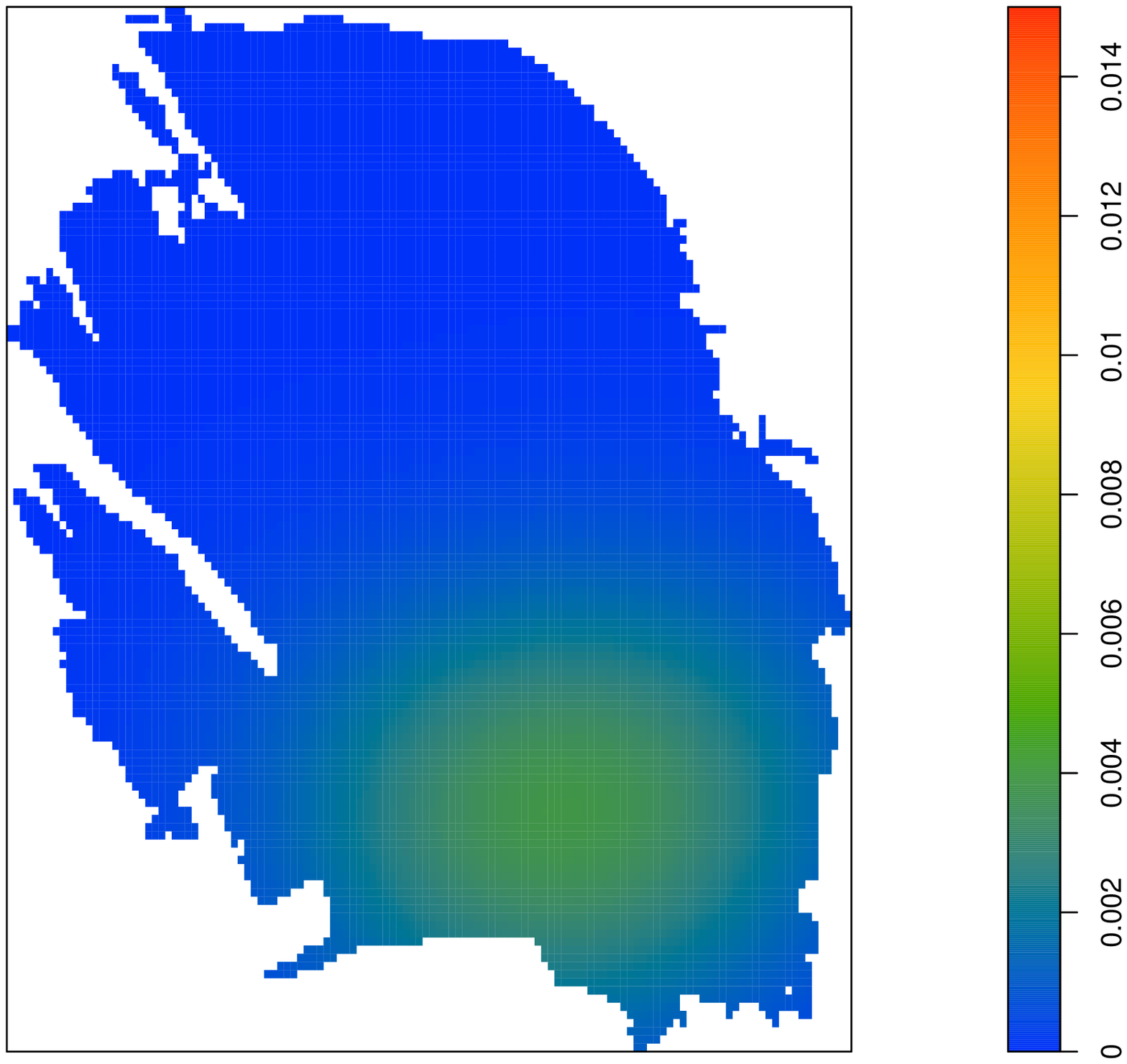}
}
\caption{Smoothed monthly gas production over January 2012 (left-most panel)
and January 2021 (right-most panel) in Nbcm per square kilometre.}
\label{F:smoothProduction}
\end{figure}

In order to smooth the production out in a mass-preserving
fashion, an adaptive kernel method similar to that proposed in
Section~\ref{S:explore} for the intensity function can be used. 
More specifically, write $V(s,t)\geq 0$ for the volume of gas produced 
at a well at $s$ in month $t$ and write $\phi$ for the well pattern.
We then use the spatial counterpart of Algorithm~\ref{A:Hamilton} 
proposed by Van Lieshout \cite{Lies21} to select 
bandwidths $h_{g,S}$ and $h_{a,S}$ and set, for $s_0 \in W_S$,
$t_0\in W_T$,
\[
\hat \lambda_V( (s_0, t_0), \phi) = 
\sum_{(s,t) \in \phi}
\frac{ V(s,t) 1\{ t_0 \in m(t) \} }{
w(s, h_{a,S}) h_{a,S}^2 \hat c(s; h_{g,S})^2 \ell(m(t))}
\kappa\left(  \frac{s_0 - s}{ h_{a,S} \hat c(s; h_{g,S})}
\right) 
\]
based on spatial kernel $\kappa$ and using 
local edge correction weights $w$, cf.\ Section~\ref{S:explore}. 
In time, $V(s,t)$ is spread evenly over the $\ell(m(t))$ days in the 
month $m(t)$ to which $t\in W_T$ belongs. This smoother is mass preserving,
as
\begin{eqnarray*}
 \int_{W_S\times W_T} \hat \lambda_V( (s, t_i), \phi) ds & = &
 \sum_{(s,t) \in \phi}
\frac{ V(s,t)}{ w(s, h_{a,S})}
 \int_{W_S}
\frac{ 1}{
 h_{a,S}^2 \hat c(s; h_{g,S})^2 }
\kappa\left(  \frac{s_0 - s}{ h_{a,S} \hat c(s; h_{g,S})}
\right) ds_0 \\
& = & 
\sum_{(s,t)\in\phi} V(s,t).
\end{eqnarray*}

For the data discussed in Section~\ref{S:gas}, Algorithm~\ref{A:Hamilton}
yields a pilot bandwidth $h_{g,S} = 7.6$ and $h_{a,S} = 6.9$. The resulting
gas production maps for January 2012, before legislation to phase out
gas extraction came into effect, and for January 2021 are shown in
Figure~\ref{F:smoothProduction}. The decrease in extracted volume is
evident. Moreover, the remaining production is located mostly in  
the southern part of the gas field.

\subsection{Regression analysis for pore pressure}
\label{S:porepressure}

From Figure~\ref{F:pressure}, it is clear that the pore pressure in
the reservoir is decreasing in time. The trend is similar for most wells, 
with some exceptions, for example because of large faults or at the
periphery of the field. Recall from Section~\ref{S:data} that our earthquake
catalogue contains tremors from January 1st, 1995 onward. If we wish to
use the pore pressure as an explanatory variable in a monitoring
model, we need their values over the same time period. Thus, our goal
in this section is to perform a regression analysis using only the
$352$ pore pressure measurements from January 1st, 1995, and later.

\begin{figure}[hbt]
\centering
\centerline{
\epsfxsize=0.45\hsize
\epsffile{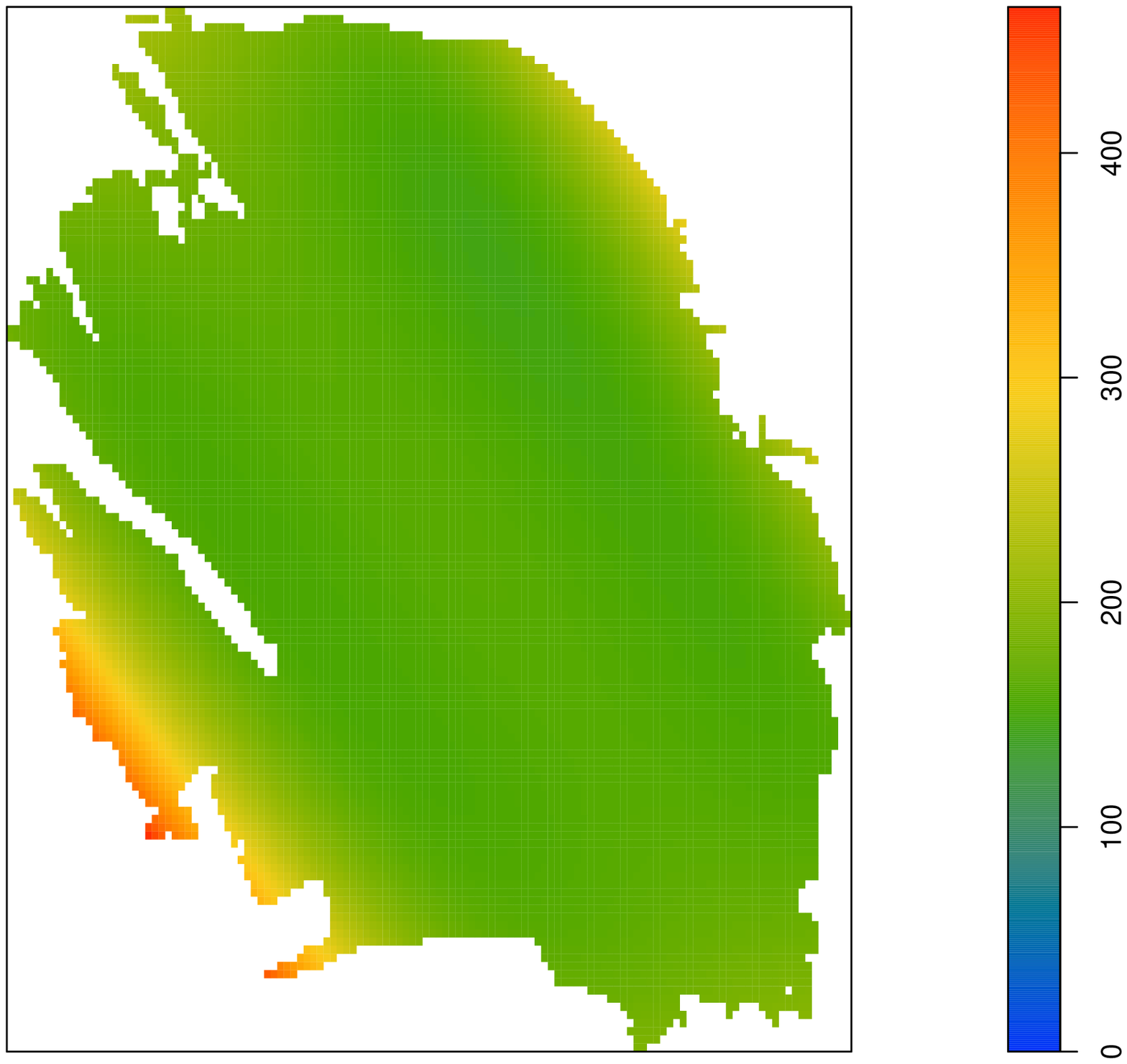}
\epsfxsize=0.45\hsize
\epsffile{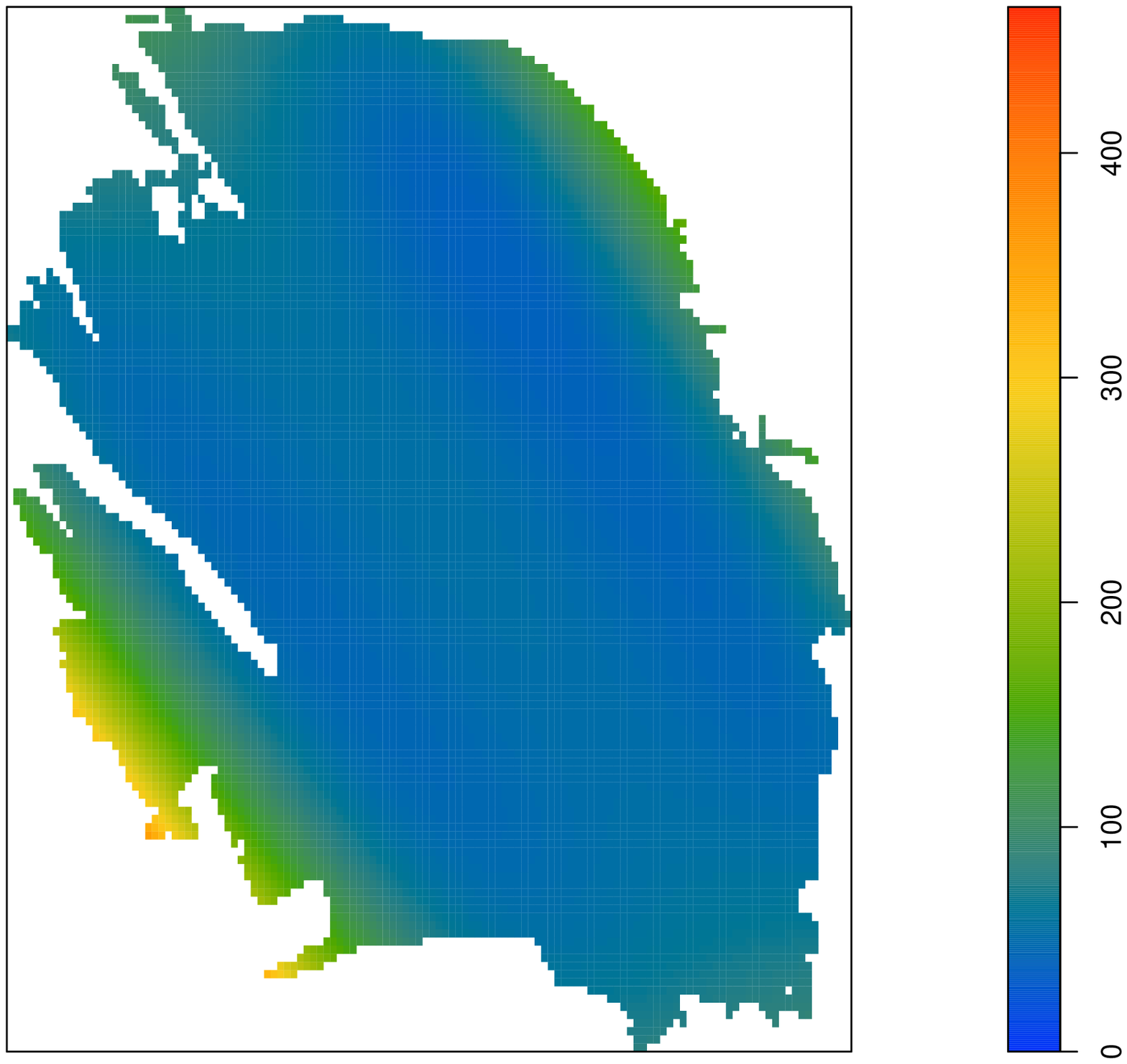}
}
\caption{Estimated pore pressure maps at midnight on January 1st in
the years 1995 (left-most
panel) and 2022 (right-most panel) in bara.}
\label{F:fitpressure95}
\end{figure}

Suppose that the observed pore pressures can be seen as realisations
of stochastic variables $X(s,t)$ that can be decomposed in a trend term
$m((s,t); \beta)$ and noise $E(s,t)$ as follows:
\[
X(s,t) = m((s,t); \beta) + E(s,t).
\]
We assume that the $E(s,t)$ are independent and normally 
distributed with mean zero and variance $\sigma^2$. For the trend
fit a polynomial in space and time,
\begin{eqnarray*}
m( (s, t); \beta ) & = & \beta_1 + \beta_2 t + \beta_3 t^2
 +  \beta_4 (s-s_0)_1 + \beta_5 (s-s_0)_2 + \beta_6 (s-s_0)_1^2 \\
& + & \beta_7 (s-s_0)_1 (s-s_0)_2 +
\beta_8 (s-s_0)_2^2 
 +  \beta_9 (s-s_0)_1^3 \\
& + & \beta_{10} (s-s_0)_1^2 (s-s_0)_2 + \beta_{11} (s-s_0)_1 (s-s_0)_2^2 +
\beta_{12} (s-s_0)_2^3   \\
& + & \beta_{13} (s-s_0)_1^4 + \beta_{14} (s-s_0)_1^3 (s-s_0)_2 + \beta_{15} (s-s_0)_1^2 (s-s_0)_2^2 \\
& + &
\beta_{16} (s-s_0)_1 (s-s_0)_2^3 + \beta_{17} (s-s_0)_2^4 
 +   \beta_{18} t (s-s_0)_1 
 +  \beta_{19} t (s-s_0)_2 \\
& + & \beta_{20} t (s-s_0)_1^2
+ \beta_{21} t (s-s_0)_1 (s-s_0)_2 + \beta_{22} t (s-s_0)_2^2 
 +  \beta_{23} t (s-s_0)_1^3 \\
& + &\beta_{24} t (s-s_0)_1^2 (s-s_0)_2
 +   \beta_{25} t (s-s_0)_1 (s-s_0)_2^2 + \beta_{26} t (s-s_0)_2^3
\end{eqnarray*}
where we centre the spatial locations at $ s_0 = (750, 5900)$ in the
UTM system (zone $31$ in units of km) and take days as the temporal
unit counting from January 1st, 1995. The parameter $\beta =
(\beta_1, \dots, \beta_{26})$ can be estimated by the least squares method.

The fitted pore pressure maps at midnight January 1st in the years
1995 and 2022 are given in Figure~\ref{F:fitpressure95}. One may note 
the elevated values in the Harkstede block in the South-West as well 
as those in the Northern border regions. 


\begin{figure}[hbt]
\centering
\centerline{
\epsfxsize=0.45\hsize
\epsffile{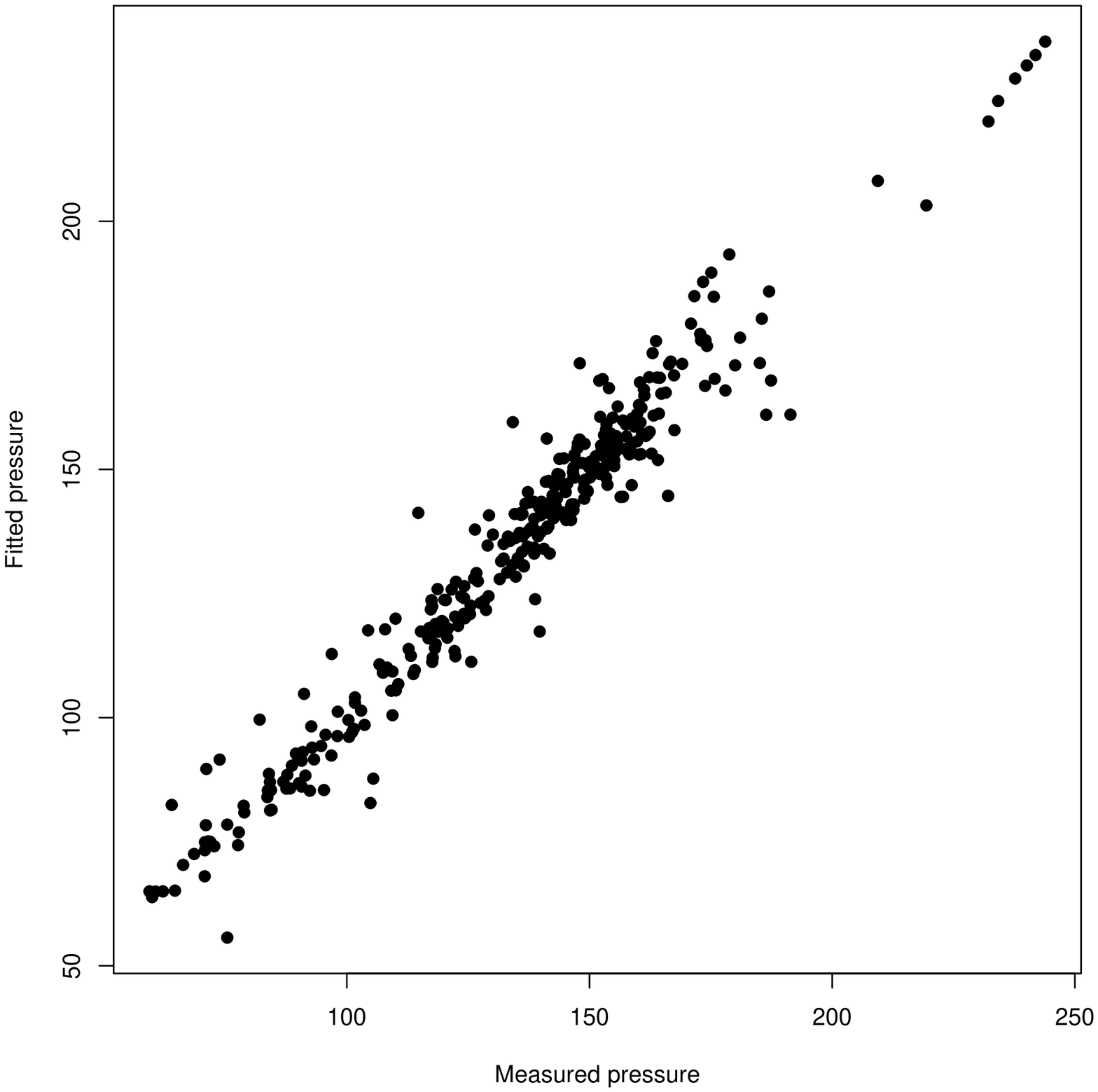}
\epsfxsize=0.45\hsize
\epsffile{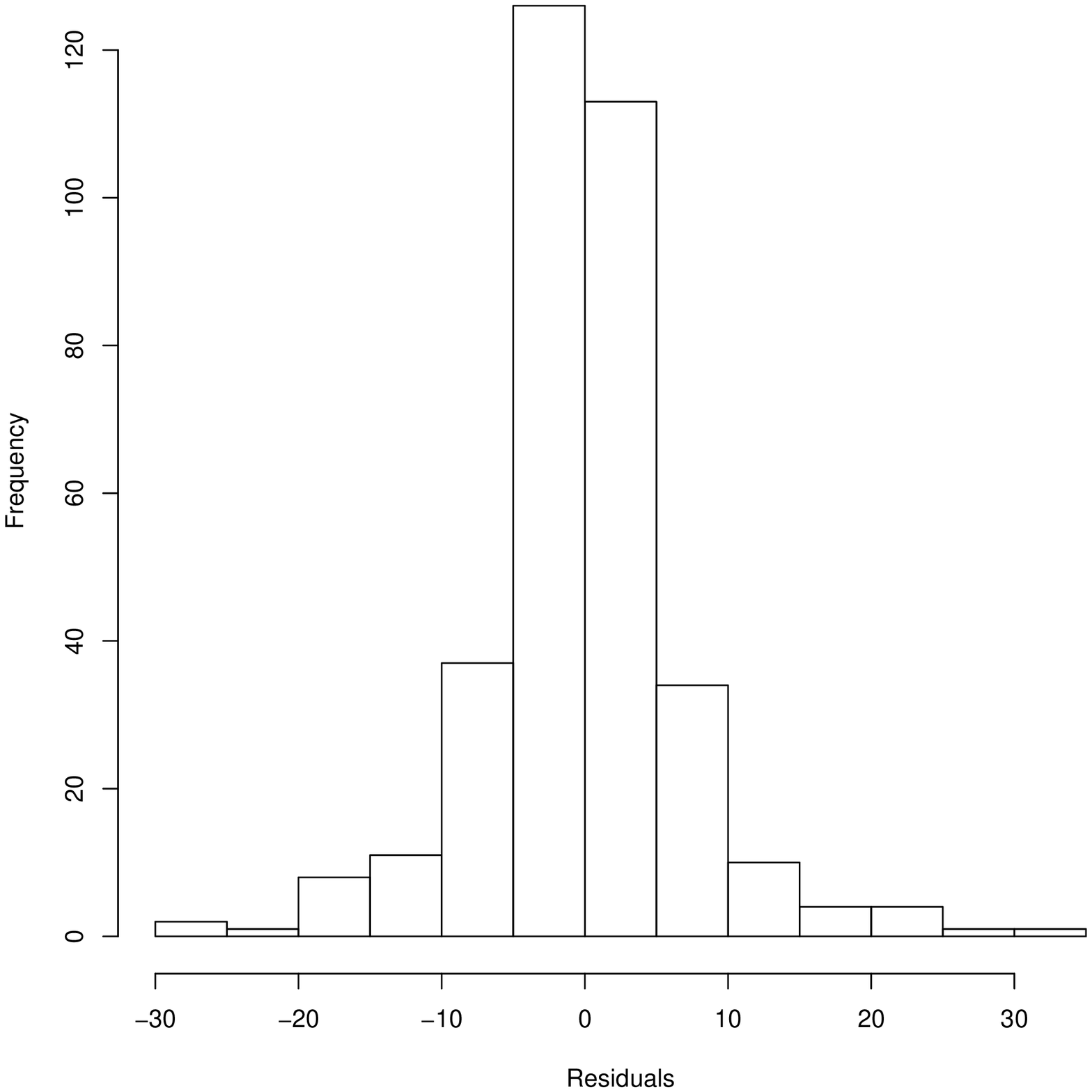}
}
\caption{Observed pore pressure values against fitted values (left-most panel) and
the histogram of residuals (right-most panel).}
\label{F:pressure95}
\end{figure}

To validate the model, the actual pore pressure measurements are plotted
against their fitted values in the left-most panel of Figure~\ref{F:pressure95}.
The graph seems reasonably close to a straight line. The histogram of the
residuals is shown in the right-most panel of Figure~\ref{F:pressure95}.
Most residuals are quite small and the histogram is centred around zero
with estimated standard deviation $\hat \sigma = 7.17$ bara.

\section{Discussion and further work}
\label{S:Cox}

In this paper, we carried out an exhaustive second order exploratory
analysis of the spatio-temporal point pattern of earthquakes recorded
in the Groningen gas field since January 1995. To do so, we needed
to develop new methodology. We proposed an adaptive kernel smoothing
technique for estimating the intensity function and suggested a practical
algorithm for selecting the spatial and temporal bandwidths. The
estimated intensity function was then plugged into state of the art
inhomogeneous summary statistics to quantify the degree of clustering
in the earthquake catalogue. We also applied our new adaptive kernel
smoothing technique to monthly gas production figures. Finally, we
performed a regression analysis on pore pressure data for the gas
field. 

In the rate-and-state models (e.g.\ Candela et al. \cite{Cand19},
Dempsey and Suckale \cite{DempSuck17} and Richter et al. \cite{Rich20})
that can be seen as the state of the art in modelling the seismic
hazard and that is being used for planning, the earthquake intensity
$\lambda$ (the rate) is assumed to be inversely proportional to a
state variable $\Gamma$, that is,
\[
\lambda(s,t)  \propto \Gamma(s,t)^{-1}, \quad (s,t)\in W_S\times W_T.
\]
The state variable $\Gamma(s,t)$ is defined by the ordinary
differential equation
\[
d\Gamma(s,t) = \alpha \left[ dt - \Gamma(s,t) dS(s,t) \right],
\]
where $S$ is the Coulomb stress with a reduced friction coefficient
and $\alpha > 0$.  According to Richter et al.\ \cite{Rich20}, Coulomb stress changes
are proportional to changes in pore pressure $X(s,t)$, that is 
$dS(s,t) = - \beta dX(s,t)$ for some scalar $\beta > 0$.
Consequently, the differential equation for the state can be written as
\[
d\Gamma(s,t) = \alpha \left[ dt + \beta \Gamma(s,t) dX(s,t) \right].
\]
Multiplying both sides by $\exp( -\alpha \beta X(s,t) )$, it follows
that 
\begin{equation} \label{odePressure}
\frac{d}{dt} \left[ \Gamma(s,t) e^{-\alpha \beta X(s,t)} \right]
= \alpha e^{-\alpha \beta X(s,t) }.
\end{equation}
The Euler discretisation reads
\[
\Gamma(s,t+\Delta) = \left( \Gamma(s,t) + \Delta \alpha \right) 
\exp\left[ \alpha \beta ( X(s,t+\Delta )  - X(s,t) ) \right],
\quad s \in W_S,
\]
upon discretising $W_T$ in time steps of length $\Delta$.

In the rate-and-state model, 
the earthquakes constitute a Poisson point process with intensity
function $\lambda$, possibly modified by a fault map. The parameters
$\alpha$ and $\beta$ and the initial state $\Gamma(s, 0) \equiv 1/\lambda_0$
are treated as unknowns and can be estimated, for example by the maximum
likelihood method.

Based on the exploratory analysis in Sections~\ref{S:explore} and
\ref{S:covariates}, the rate-and-state model can be criticised on several
points. As we saw in  Section~\ref{S:explore}, the earthquake pattern
exhibits clustering. Since by definition the points in any Poisson point
process do not interact with one another, the apparent clustering of
earthquakes cannot be described by the rate-and-state model. Secondly,
the pressure values are assumed to be known everywhere, in practice by
interpolation of the measurements. Proceeding in this way, the uncertainty
in the interpolations is ignored. It would be better to treat the $X(s,t)$ as
a random field. Lastly, the varying gas extraction is not taken into account.

Based on the above considerations, we propose the following model.
Set $\Gamma(s,0)  =  1 / \lambda_0$ and iterate
\[
\Gamma(s,t+\Delta)  =  \left( \Gamma(s,t) + \alpha \Delta \right) 
  \exp\left[\alpha \beta ( m(s,t+\Delta) - m(s,t) ) \right] 
  \exp\left[ \alpha \beta ( E(s, t+ \Delta) - E(s,t) )\right]
\]
where $m$ and $E$ are as in Section~\ref{S:porepressure}. 
The gas production and random effects, if any, may be included
in the following way. Let $\Psi$ be a Cox process (cf.\ Chiu et al.\
\cite{SKM}) on 
$W_S \times W_T$ with driving random measure
\[
\Lambda(s,t) = 
\frac{\exp\left[ 
   \theta_1 + \theta_2 \tilde V(s,t) + U(s,t)
\right]}{ \Gamma(s,t)}
\]
where $\tilde V(s,t)$ is the gas extracted at $s$ during the year
preceeding time $t$ and $U(s,t)$ is a correlated Gaussian field 
that accounts for random effects. Monitoring can then be based 
on the posterior distribution of $\Lambda$ or, equivalently $\Gamma$
and $U$, given the recorded earthquakes. The implementation requires 
careful use of Markov chain Monte Carlo techniques and is the topic 
of our ongoing research.

\section*{Acknowledgements}

This research was funded by the Dutch Research Council NWO through
their DEEPNL programme (grant number DEEP.NL.2018.033).
We are grateful to Professor Van Dinther for expert advice and to
Mr Rob van Eijs for providing us with data on gas production.


\begin{thebibliography}{99}

\bibitem{Abra82a}
Abramson, I.S. (1982).
On bandwidth variation in kernel estimates -- A square root law.
{\em The Annals of Statistics\/} 10:1217--1223.

\bibitem{BakiLies22}
Baki, Z., Lieshout, M.N.M.~van (2022).
The influence of gas production on seismicity in the Groningen field.
{\em Proceedings of the 10th International Workshop on Spatio-Temporal
Modelling METMA X\/}, pp.~163--167.

\bibitem{Bour18}
Bourne, S.J., Oates, S.J., Van Elk, J.\ (2018).
The exponential rise of induced seismicity with increasing stress levels
in the Groningen gas field and its implications for controlling seismic risk. 
{\em Geophysicial Journal International\/} 213:1693--1700.

\bibitem{Cand19}
Candela, T.\ et al. (2019).
Depletion-induced seismicity at the Groningen gas field: Coulomb
rate-and-state models including differential compaction effect.
{\em Journal of Geophysical Research: Solid Earth\/} 124:7081--7104.

\bibitem{ChacDuon18}
Chac\'on, J.E., Duong, T. (2018).
{\em Multivariate Kernel Smoothing and its Applications}.
CRC Press.

\bibitem{SKM}
Chiu, S.N., Stoyan, D., Kendall, W.S., Mecke, J. (2013).
{\em Stochastic Geometry and its Applications}.
Wiley, 3rd ed.

\bibitem{CronLies15}
Cronie, O., Lieshout, M.N.M.~van (2015).
A $J$-function for inhomogeneous spatio-temporal point processes.
{\em Scandinavian Journal of Statistics} 42:562--579.

\bibitem{CronLies18}
Cronie, O., Lieshout, M.N.M.~van (2018).
A non-model based approach to bandwidth selection for kernel
estimators of spatial intensity functions.
{\em Biometrika\/} 105:455--462.

\bibitem{Davi18}
Davies, T.M., Flynn, C.R., Hazelton, M.L. (2018).
On the utility of asymptotic bandwidth selectors for 
spatially adaptive kernel density estimation.
{\em Statistics and Probabability Letters\/} 138:75--81.

\bibitem{DempSuck17}
Dempsey, D., Suckale, J. (2017).
Physics-based forecasting of induced seismicity at Groningen
gas field, the Netherlands.
{\em Geophysical Research Letters\/} 22:7773--7782.

\bibitem{Digg85}
Diggle, P.\ (1985).
A kernel method for smoothing point process data.
{\em Applied Statistics\/} 34:138--147.

\bibitem{Dost12}
Dost, B., Goutbeek, F., Van Eck, T., Kraaijpoel, D.\ (2012).
Monitoring induced seismicity in the North of the Netherlands:
Status report 2010. 
{\em Scientific Report KNMI} WR 2012--03.

\bibitem{GabrDigg09}
Gabriel, E., Diggle, P. J. (2009).
Second-order analysis of inhomogeneous spatio-temporal point process data.
{\em Statistica Neerlandica\/} 63:43--51

\bibitem{Geer14}
Geerdink, E.\ (2014).
Modeling the induced earthquakes in Groningen as a Poisson process using GLM and GAM.
BSc thesis, University of Groningen

\bibitem{Hett17}
Hettema, M.H.H., Jaarsma, B., Schroot, B.M., Van Yperen, G.C.N.\ (2017).
An empirical relationship for the seismic activity rate of the Groningen gas field.
{\em Netherlands Journal of Geosciences\/} 96:149--161.

\bibitem{Hove15}
Hove, E.~van, Van Lingen, R., Riemens, S.\ (2015).
Ge\"induceerde aardbevingen in gasveld Groningen. Een statistische analyse. 
BSc thesis, University of Twente.

\bibitem{Lies11}
Lieshout, M. N. M.~van (2011).
A $J$-function for inhomogeneous point processes.
{\em Statistica Neerlandica\/}  65:183--201

\bibitem{Lies12}
Lieshout, M.N.M.~van (2012).
On estimation of the intensity function of a point process.
{\em Methodology and Computing in Applied Probability\/} 14:567--578.

\bibitem{Lies20}
Lieshout, M.N.M.~van (2020).
Infill asymptotics and bandwidth selection for kernel estimators
of spatial intensity functions.
{\em Methodology and Computing in Applied Probability\/} 22:995--1008.

\bibitem{Lies21}
Lieshout, M.N.M.van (2021).
Infill asymptotics for adaptive kernel estimators of spatial intensity.
{\em Australian and New Zealand Journal of Statistics\/} 63:159--181, 2021.

\bibitem{Lo17}
Lo, P.H. (2017).
An iterative plug-in algorithm for optimal bandwidth
selection in kernel intensity estimation for spatial data.
PhD Thesis, Technical University of Kaiserslautern.
  
\bibitem{Ogat88}
Ogata, Y.\ (1988).
Statistical models for earthquake occurrences and residual analysis for point processes.
{\em Journal of the American Statistical Association\/} 83-401:9--27.

\bibitem{Post21}
Post, R.A.J. et al. (2021).
Interevent-time distribution and aftershock frequency in non-stationary 
induced seismicity. 
{\em Scientific Reports\/} 11, 3540.

\bibitem{Rich20}
Richter, G., Hainzl, S., Dahm, T., Z\H{o}ller, G. (2020).
Stress-based statistical modeling of the induced seismicity at the
Groningen gas field, The Netherlands.
{\em Environmental Earth Sciences\/} 79, 252.
  
\bibitem{Sija17}
Sijacic, D., Pijpers, F., Nepveu, M., Van Thienen--Visser, K.\ (2017).
Statistical evidence on the effect of production changes on induced seismicity.
{\em Netherlands Journal of Geosciences\/} 96:27--38.

\bibitem{Tram22}
Trampert J., Benzi R., Toschi F. (2022).
Implications of the statistics of seismicity recorded within the
Groningen gas field.
{\em Netherlands Journal of Geosciences\/} 101, to appear.

\bibitem{Vlek19}
Vlek, C. (2019).
Rise and reduction of induced earthquakes in the Groningen
gas field, 1991--2018: Statistical trends, social impacts, and
policy change.
{\em Environmental Earth Sciences\/} 78:1--14.

\end{thebibliography}
\end{document}